\documentclass[nolinenum]{jfp}

\usepackage{graphicx}
\usepackage{fancyvrb}
\usepackage{mathpartir}
\usepackage{mathtools}
\usepackage{bussproofs}
\usepackage{multicol}

\usepackage{subcaption}
\usepackage{caption}
\usepackage{tikz}
\usepackage{tikz-cd}
\usepackage{csvsimple,booktabs, siunitx, array}
\usepackage{sansmath}
\usepackage{stackengine}
\usepackage{pict2e}
\usepackage{xspace}
\usepackage{proof}

\newcommand{\ponders}[3]{\ignorespaces}
\newcommand{\TODO}[1]{}

\makeatletter
\@ifundefined{lhs2tex.lhs2tex.sty.read}%
  {\@namedef{lhs2tex.lhs2tex.sty.read}{}%
   \newcommand\SkipToFmtEnd{}%
   \newcommand\EndFmtInput{}%
   \long\def\SkipToFmtEnd#1\EndFmtInput{}%
  }\SkipToFmtEnd

\newcommand\ReadOnlyOnce[1]{\@ifundefined{#1}{\@namedef{#1}{}}\SkipToFmtEnd}
\usepackage{amstext}
\usepackage{amssymb}
\usepackage{stmaryrd}
\DeclareFontFamily{OT1}{cmtex}{}
\DeclareFontShape{OT1}{cmtex}{m}{n}
  {<5><6><7><8>cmtex8
   <9>cmtex9
   <10><10.95><12><14.4><17.28><20.74><24.88>cmtex10}{}
\DeclareFontShape{OT1}{cmtex}{m}{it}
  {<-> ssub * cmtt/m/it}{}

\DeclareFontShape{OT1}{cmtt}{bx}{n}
  {<5><6><7><8>cmtt8
   <9>cmbtt9
   <10><10.95><12><14.4><17.28><20.74><24.88>cmbtt10}{}
\DeclareFontShape{OT1}{cmtex}{bx}{n}
  {<-> ssub * cmtt/bx/n}{}

\newcommand{\Conid}[1]{\mathit{#1}}
\newcommand{\Varid}[1]{\mathit{#1}}
\newcommand{\anonymous}{\kern0.06em \vbox{\hrule\@width.5em}}

\newcommand{\bind}{\mathbin{>\!\!\!>\mkern-6.7mu=}}

\usepackage{polytable}

\@ifundefined{mathindent}%
  {\newdimen\mathindent\mathindent\leftmargini}%
  {}%

\def\resethooks{%
  \global\let\SaveRestoreHook\empty
  \global\let\ColumnHook\empty}
\newcommand*{\savecolumns}[1][default]%
  {\g@addto@macro\SaveRestoreHook{\savecolumns[#1]}}
\newcommand*{\restorecolumns}[1][default]%
  {\g@addto@macro\SaveRestoreHook{\restorecolumns[#1]}}
\newcommand*{\aligncolumn}[2]%
  {\g@addto@macro\ColumnHook{\column{#1}{#2}}}

\resethooks

\newcommand{\onelinecommentchars}{\quad-{}- }
\newcommand{\commentbeginchars}{\enskip\{-}
\newcommand{\commentendchars}{-\}\enskip}

\newcommand{\visiblecomments}{%
  \let\onelinecomment=\onelinecommentchars
  \let\commentbegin=\commentbeginchars
  \let\commentend=\commentendchars}

\newcommand{\invisiblecomments}{%
  \let\onelinecomment=\empty
  \let\commentbegin=\empty
  \let\commentend=\empty}

\visiblecomments

\newlength{\blanklineskip}
\newcommand{\hsindent}[1]{\quad}%
\let\hspre\empty
\let\hspost\empty

\EndFmtInput
\makeatother
\ReadOnlyOnce{polycode.fmt}%
\makeatletter

\newcommand{\hsnewpar}[1]%
  {{\parskip=0pt\parindent=0pt\par\vskip #1\noindent}}

\newcommand{\hscodestyle}{}

\newcommand{\sethscode}[1]%
  {\expandafter\let\expandafter\hscode\csname #1\endcsname
   \expandafter\let\expandafter\endhscode\csname end#1\endcsname}

  {\par\noindent
   \advance\leftskip\mathindent
   \hscodestyle
   \let\\=\@normalcr
   \let\hspre\(\let\hspost\)%
   \pboxed}%
  {\endpboxed\)%
   \par\noindent
   \ignorespacesafterend}

  {\hsnewpar\abovedisplayskip
   \advance\leftskip\mathindent
   \hscodestyle
   \let\hspre\(\let\hspost\)%
   \pboxed}%
  {\endpboxed%
   \hsnewpar\belowdisplayskip
   \ignorespacesafterend}

  {\hsnewpar\abovedisplayskip
   \advance\leftskip\mathindent
   \hscodestyle
   \let\\=\@normalcr
   \(\pboxed}%
  {\endpboxed\)%
   \hsnewpar\belowdisplayskip
   \ignorespacesafterend}

\newcommand{\plainhs}{\sethscode{plainhscode}}

\plainhs

  {\hsnewpar\abovedisplayskip
   \advance\leftskip\mathindent
   \hscodestyle
   \let\\=\@normalcr
   \(\parray}%
  {\endparray\)%
   \hsnewpar\belowdisplayskip
   \ignorespacesafterend}

  {\parray}{\endparray}

  {\(\parray}{\endparray\)}

\def\codeframewidth{\arrayrulewidth}
\RequirePackage{calc}

  {\parskip=\abovedisplayskip\par\noindent
   \hscodestyle
   \arrayrulewidth=\codeframewidth
   \tabular{@{}|p{\linewidth-2\arraycolsep-2\arrayrulewidth-2pt}|@{}}%
   \hline\framedhslinecorrect\\{-1.5ex}%
   \let\endoflinesave=\\
   \let\\=\@normalcr
   \(\pboxed}%
  {\endpboxed\)%
   \framedhslinecorrect\endoflinesave{.5ex}\hline
   \endtabular
   \parskip=\belowdisplayskip\par\noindent
   \ignorespacesafterend}

\newcommand{\framedhslinecorrect}[2]%
  {#1[#2]}

  {\(\def\column##1##2{}%
   \let\>\undefined\let\<\undefined\let\\\undefined
   \newcommand\>[1][]{}\newcommand\<[1][]{}\newcommand\\[1][]{}%
   \def\fromto##1##2##3{##3}%
   }{\) }%

  {\let\orighscode=\hscode
   \let\origendhscode=\endhscode
   \def\endhscode{\def\hscode{\endgroup\def\@currenvir{hscode}\\}\begingroup}
   \orighscode\def\hscode{\endgroup\def\@currenvir{hscode}}}%
  {\origendhscode
   \global\let\hscode=\orighscode
   \global\let\endhscode=\origendhscode}%

\makeatother
\EndFmtInput
\ReadOnlyOnce{forall.fmt}%
\makeatletter

\let\HaskellResetHook\empty
\newcommand*{\AtHaskellReset}[1]{%
  \g@addto@macro\HaskellResetHook{#1}}
\newcommand*{\HaskellReset}{\HaskellResetHook}

\newcommand\hsforall{\global\let\hsdot=\hsperiodonce}
\newcommand*\hsperiodonce[2]{#2\global\let\hsdot=\hscompose}
\newcommand*\hscompose[2]{#1}

\AtHaskellReset{\global\let\hsdot=\hscompose}

\HaskellReset

\makeatother
\EndFmtInput

\DefineVerbatimEnvironment%
  {core}{Verbatim}
  {xleftmargin=\mathindent}

\newsavebox{\lbansbox}
\newcommand{\lbansmacro}{%
\begin{tikzpicture}[baseline=0.45ex,xscale=0.006em,yscale=0.012ex]%
\draw[solid,join=round] (1.8,6) -- (1.5,5.9) to[out=-120, in=90] (0.7,3) to[out=-90, in=120] (1.5,0.1) -- (1.8,0) -- cycle;%
\end{tikzpicture}}
\savebox{\lbansbox}{\lbansmacro}
\newcommand{\lbans}{\mathopen{\usebox{\lbansbox}\mspace{1mu}}}

\newsavebox{\rbansbox}
\newcommand{\rbansmacro}{%
\begin{tikzpicture}[baseline=0.45ex,xscale=-0.006em,yscale=0.012ex]%
\draw[solid,join=round] (1.8,6) -- (1.5,5.9) to[out=-120, in=90] (0.7,3) to[out=-90, in=120] (1.5,0.1) -- (1.8,0) -- cycle;%
\end{tikzpicture}}
\savebox{\rbansbox}{\rbansmacro}
\newcommand{\rbans}{\mathclose{\mspace{1mu}\usebox{\rbansbox}}}

\newsavebox{\llensbox}
\newcommand{\llensmacro}{%
\begin{tikzpicture}[baseline=0.45ex,xscale=0.006em,yscale=0.012ex]%
\draw[solid,join=round] (1.4,0) -- (0,0) -- (0,6) -- (1.4,6) -- (1.5,5.9) to[out=-120, in=90] (0.7,3) to[out=-90, in=120] (1.5,0.1) -- cycle;%
\end{tikzpicture}}
\savebox{\llensbox}{\llensmacro}
\newcommand{\llens}{\mathopen{\usebox{\llensbox}\mspace{1mu}}}

\newsavebox{\rlensbox}
\newcommand{\rlensmacro}{%
\begin{tikzpicture}[baseline=0.45ex,xscale=-0.006em,yscale=0.012ex]%
\draw[solid,join=round] (1.4,0) -- (0,0) -- (0,6) -- (1.4,6) -- (1.5,5.9) to[out=-120, in=90] (0.7,3) to[out=-90, in=120] (1.5,0.1) -- cycle;%
\end{tikzpicture}}
\savebox{\rlensbox}{\rlensmacro}
\newcommand{\rlens}{\mathclose{\mspace{1mu}\usebox{\rlensbox}}}

\newcommand{\catamor}[1]{\lbans\, #1\, \rbans}

\newcommand{\ana}[1]{\llens #1 \rlens}

\newcommand{\bigstep}{\Downarrow}
\newcommand{\Nat}{\mathbb{N}}

\newcommand{\den}[1]{\llbracket #1 \rrbracket}

\newcommand{\operator}[1]{\texttt{#1}\xspace}
\newcommand{\iso}{\cong}

\newcommand{\eqiff}{\stackrel{\triangle}{\iff}}

\newcommand{\Id}{\text{Id}}

\newcommand{\floorCoPointed}[1]{\lfloor #1 \rfloor_{\times}}
\newcommand{\ceilingCoPointed}[1]{\lceil #1 \rceil_{\times}}

\newcommand{\floorFree}[1]{\lfloor #1 \rfloor_{*}}
\newcommand{\ceilingFree}[1]{\lceil #1 \rceil_{*}}

\newcommand{\copointed}[1]{#1_{\times}}
\newcommand{\free}[1]{#1^{*}}

\newcommand{\finpow}[1]{\mathcal{P}_{\text{fin}}(#1)}

\newcommand{\Act}{\textsf{Act}}

\newcommand{\nil}{\operator{nil}}
\newcommand{\rep}{\operator{rep}}
\newcommand{\restr}{\backslash}
\newcommand{\Chan}{\ensuremath{\Conid{Chan}}}
\newcommand{\Val}{\mathbb{V}}
\newcommand{\pow}[1]{\mathcal{P}(#1)}

\newenvironment{codepage}[1][.9\linewidth]
  { 
    \begin{minipage}{#1}
    \vspace{-\abovedisplayskip} }
  { \vspace{-\belowdisplayskip}\vspace{-0.7\baselineskip}
    \end{minipage} }

\theoremstyle{plain}%

\theoremstyle{definition}

\newtheorem{example}{Example}[section]
\theoremstyle{remark}

\begin{document}

\journaltitle{JFP}
\cpr{Cambridge University Press}
\doival{10.1017/xxxxx}

\lefttitle{Semantics {\`a} la carte}
\righttitle{Journal of Functional Programming}

\totalpg{\pageref{lastpage01}}
\jnlDoiYr{2022}

\begin{authgrp}
\author{Marco Paviotti}
\affiliation{University of Kent,\\
        (\email{m.paviotti@kent.ac.uk})}
\author{Nicolas Wu}
\affiliation{Imperial College London,\\
        (\email{n.wu@imperial.ac.uk})}
\end{authgrp}

\title{Full Abstraction for Free}

\begin{abstract}
Structured recursion schemes such as folds and unfolds have been widely used for
structuring both functional programs and program semantics. In this context, it
has been customary to implement denotational semantics as folds over an
inductive data type to ensure termination and compositionality.  Separately,
operational models can be given by unfolds, and naturally not all operational
models coincide with a given denotational semantics in a meaningful way.

To ensure these semantics are coherent it is important to consider the
property of full abstraction which relates the denotational and the operational
model. In this paper, we show how to engineer a compositional semantics such that full
abstraction comes for free. We do this by using \emph{distributive laws} from
which we generate both the operational and the denotational model. The
distributive laws ensure the semantics are fully abstract at the type level,
thus relieving the programmer from the burden of the proofs.

\end{abstract}

\maketitle[F]

\newcommand{\then}{\text{\xspace then \xspace }}
\newcommand{\fa}{\text{\xspace for all \xspace }}

\section{Introduction}
\label{sec:intro}

For a long time structured recursion schemes have been used  to ensure that
functions are well defined~\citep{MeijerFP91}. In particular, folds have been
used to ensure termination while unfolds have been used to ensure productivity
of recursively defined functions. A perhaps less known fact about recursion
operators is that they provide additional semantic properties when used in the
context of programming languages and semantics. \citet{Hutton98} popularized the
idea that folds can be used to define the denotational semantics, and unfolds
can be used to define the operational semantics of such languages.

Given that folds and unfolds provide different semantics, a natural question is
how they relate to one another. 
This question is answered with the idea of \emph{full abstraction} which is the
statement that syntactically different programs that behave \emph{operationally}
the same way in any context should be regarded as exactly the same entity
\emph{denotationally}.%

To explain this point we define the grammar of Hutton's razor as an inductive
set (left) and implement it as a data type in Haskell (right):

{
\setlength{\mathindent}{0cm}
\begin{minipage}{0.5\linewidth}
  \begin{center}$L := \Nat \mid L + L$\end{center}
\end{minipage}
\begin{minipage}{0.5\linewidth}
\begin{hscode}\SaveRestoreHook
\column{B}{@{}>{\hspre}l<{\hspost}@{}}%
\column{E}{@{}>{\hspre}l<{\hspost}@{}}%
\>[B]{}\mathbf{data}\;\Conid{L}\mathrel{=}\Conid{Val}\;\Conid{Nat}\mid \Conid{Add}\;\Conid{L}\;\Conid{L}{}\<[E]%
\ColumnHook
\end{hscode}\resethooks
\end{minipage}
}
\noindent
where \ensuremath{\Conid{Nat}} is the natural numbers type. The semantics of this language can be
defined by folding the structure of the syntax into the structure of a semantic
domain, given by the base case \ensuremath{\Varid{n}\mathbin{::}\Conid{Nat}} and the binary operation \ensuremath{\Conid{Add}\mathbin{::}\Conid{Nat}\to \Conid{Nat}}:
\begin{hscode}\SaveRestoreHook
\column{B}{@{}>{\hspre}l<{\hspost}@{}}%
\column{21}{@{}>{\hspre}l<{\hspost}@{}}%
\column{E}{@{}>{\hspre}l<{\hspost}@{}}%
\>[B]{}\llbracket\cdot \rrbracket\mathbin{::}\Conid{L}\to \Conid{Nat}{}\<[E]%
\\
\>[B]{}\llbracket\Conid{Val}\;\Varid{n}\rrbracket{}\<[21]%
\>[21]{}\mathrel{=}\Varid{n}{}\<[E]%
\\
\>[B]{}\llbracket\Conid{Add}\;\Varid{t}_{1}\;\Varid{t}_{2}\rrbracket{}\<[21]%
\>[21]{}\mathrel{=}\llbracket\Varid{t}_{1}\rrbracket\mathbin{+}\llbracket\Varid{t}_{2}\rrbracket{}\<[E]%
\ColumnHook
\end{hscode}\resethooks

\noindent
This kind of semantics is often called a \emph{denotational} semantics, where
the meaning of a term is given in terms of subterms and in this sense
denotational semantics are modular interpretations of syntax into a given domain.
As it happens, this is precisely how a generic fold operates: it takes a
function that handles a  base case and another for the inductive case, together
known as an \emph{algebra}, and uses these to reduce a data structure to a final
value. Folds over syntax exactly correspond to the structure of denotational
semantics.

However, we could have defined the semantics in many other ways that correspond
to folds as as well. For example, by sending every program to \ensuremath{()}, i.e. \ensuremath{\llbracket\anonymous \rrbracket\mathrel{=}()}, sending every program to itself, i.e. \ensuremath{\llbracket\cdot \rrbracket\mathrel{=}\Varid{id}}.
A perhaps more extreme example would be to interpret the syntactic constructor
\ensuremath{\Conid{Add}} as the semantics multiplication.

Given the multitude of different semantics that can be given, a good question is
how to provide a means of arbitrating between them. While addition of numbers
into \ensuremath{\Conid{Nat}} seems reasonable, the other three are somewhat extreme and
unsatisfactory.%

However, the only way of arbitrating between a good semantic model and a bad one
is to define precisely what programs are supposed to ``do'', for example, by giving an
\emph{operational} model for the language.  Here, rules are used
to define a reduction relation, written $\ensuremath{\Varid{t}} \to \ensuremath{\Varid{t'}}$, for programs \ensuremath{\Varid{t}} and
\ensuremath{\Varid{t'}}, where for each rule, reductions below a line are permissible given
reductions above the line:
\newcommand{\reduceto}[1]{\overset{#1}{\rightsquigarrow}}
\begin{mathpar}
  \inferrule{ }{\ensuremath{\Conid{Add}\;(\Conid{Val}\;\Varid{m})\;(\Conid{Val}\;\Varid{n})} \reduceto{} \ensuremath{\Conid{Val}\;(\Varid{m}\mathbin{+}\Varid{n})}} \quad
  \inferrule{\ensuremath{\Varid{t}_{1}} \reduceto{} \ensuremath{\Varid{t}_{1}'}}{\ensuremath{\Conid{Add}\;\Varid{t}_{1}\;\Varid{t}_{2}} \reduceto{} \ensuremath{\Conid{Add}\;\Varid{t}_{1}'\;\Varid{t}_{2}}} \quad
  \inferrule{\ensuremath{\Varid{t}_{2}} \reduceto{} \ensuremath{\Varid{t}_{2}'}}{\ensuremath{\Conid{Add}\;\Varid{t}_{1}\;\Varid{t}_{2}} \reduceto{} \ensuremath{\Conid{Add}\;\Varid{t}_{1}\;\Varid{t}_{2}'}}
\end{mathpar}

Once the operational model has been defined we can state the properties we want
from a denotational model. In particular, a good denotational model has to be
\emph{sound}, which means it has to be agnostic to step reductions and,
moreover, it has to be at least \emph{computationally adequate}, that is, if two
programs denote the same object in the model they should be operationally
indistinguishable. Furthermore, if the model equates exactly the programs that
ought to be indistinguishable the semantics is called \emph{fully abstract}.
Thus a good denotational semantics should be related in some way to the
operational semantics.

The problem is that the existing recursion schemes proposed to give semantic
interpretations, like folds, do not have any restrictions on the kind of
algebras the user can provide. In other words, folds operators do not preserve
nor reflect the additional information provided by the operational semantics.

In coalgebraic systems, we can associate an operational semantics to the syntax
tree structure at hand by using a \emph{coalgebra}, that is a function which
takes a program as input and outputs the target program along with some
information about the behavior produced. For example, the operational semantics
for the language defined above can be equivalently defined as a function \ensuremath{\Varid{opsem}\mathbin{::}\Conid{L}\to [\mskip1.5mu \Conid{L}\mskip1.5mu]}\footnote[1]{Where, for simplicity, sets are modelled as lists.}
where values are sent to the empty set of programs, programs of the form \ensuremath{\Conid{Add}\;(\Conid{Val}\;\Varid{n})\;(\Conid{Val}\;\Varid{m})} are sent to \ensuremath{\Conid{Val}\;(\Varid{n}\mathbin{+}\Varid{m})} and programs of the form \ensuremath{\Conid{Add}\;\Varid{t}_{1}\;\Varid{t}_{2}}
are sent to programs of the form \ensuremath{\Conid{Add}\;\Varid{t}_{1}'\;\Varid{t}_{2}} and \ensuremath{\Conid{Add}\;\Varid{t}_{1}\;\Varid{t}_{2}'} whenever \ensuremath{\Varid{t}_{1}}
reduces to \ensuremath{\Varid{t}_{1}'} and similarly for \ensuremath{\Varid{t}_{2}}.

Now that we have the operational information we need to ensure this is somehow
respected when interpreting the syntax. As we stated earlier, the fold operator
is not suitable for this job, but the \emph{unfold} operator is.  In particular,
the unfold is a function that given a program produces the ``trace'' generated
by ``unfolding'' or ``running'' the operational model on that program. In other
words, the unfold operator is the unique \emph{fully abstract} translation from
syntax to the domain of traces if, and only if, the unfolding of two programs is
equal whenever these programs \emph{behave the same operationally}.

In this paper we show how to make use of \emph{distributive laws} to ensure that
the interpretation function from the syntax to the semantic domain is both a
fold (is compositional) and an unfold (is fully abstract). The theory behind
this approach has been known for a long time~\citep{RuttenT93, TuriP97,
JacobsR12, Rutten00}, and many have used this theory for structuring
functional programming~\citep{HinzeJ11, JaskelioffGH11}. However, none of these
works focusses explicitly on the use of recursion operators for semantic
properties.

In this paper, we do this by giving a fresh account on the Hutton's razor and
Milner's CCS. The latter, in particular, gives raise essentially to a (correct)
Domain Specific Language (DSL) for stream programming in Haskell. Our work can
be seen as extending the work of \citet{Hutton98} which expressed they idea of
denotational semantics as a fold and of operational semantics as an unfold.
However, these two do not necessarily coincide and, in the case of CCS, the
correspondence needs a pen and paper proof.

In our approach the parametric type structure of distributive laws is used to
guide the programmer towards a fully abstract semantics for free. We found that,
while CCS semantics naturally fits in the context of the distributive laws,
Hutton's arithmetic language needs a slight tweak.
As it is customary~\citep{DanielssonHJG06}, we implement the theory from the
literature in the total fragment of Haskell. In particular, we implement data
types using recursive types and we use the appropriate recursion scheme to keep
initial algebras and final coalgebras distinct, and, finally, we only use
predicative polymorphism, so the Haskell programs in this paper can be
interpreted in the category of sets.

We start by providing background material on folds and unfolds
(\autoref{sec:folds-unfolds}) and then we use these to implement a fully
abstract semantics for a simple arithmentic language for streams
(\autoref{sec:simple-laws}) and an arithmentic language with non-determinism
(\autoref{sec:copointed}). As a more involved example, we show how give a
full abstract semantics of Milner's value-passing CCS (\autoref{sec:free-laws}).
Finally, we conclude by listing the related work (\autoref{sec:relatedwork}) and
discussion (\autoref{sec:conclusions}).

\newpage

\section{Background: Folds and Unfolds}
\label{sec:folds-unfolds}\label{sec:background}
Folds and unfolds are the most basic patterns of recursion. The former are used
to ensure termination of recursive definitions and the latter are used to ensure
definitions of infinite data structures be productive.
When folds are to structure semantics they provide a modular denotational
semantics, as popularised by~\cite{Hutton98}. Similarly, unfolds provide a fully
abstract interpretation by construction, but this fact is far less explored in
the functional programming literature. This section provides a summary the
basics of these two patterns of recursion. In Section~\ref{sec:folds} we
introduce folds. In Section~\ref{sec:unfolds} we introduce unfolds and explain
why these are fully abstract interpretations over transition systems.

\subsection{Folds}
\label{sec:folds}
A fold is a recursive operator that recurses over an inductive data structure
and through the use of a given combining operator reassembles the results of the
recursively transformed subparts of the inductive structure.

Folds use the notion of \ensuremath{\Conid{Functor}} to describe the structure of a data type and
define an inductive data type by the least fixed-point of this functor.  In the
code below, \ensuremath{\Sigma} is a variable which we will use to pass the functor as an
argument to \ensuremath{\mu\cdot }. This in turn will compute the least fixed-point of
\ensuremath{\Sigma} providing the constructor \ensuremath{\Conid{In}} and deconstructor \ensuremath{\Varid{in^\circ}} which witness the fact that \ensuremath{\mu\Sigma} is
isomorphic to \ensuremath{\Sigma\;(\mu\Sigma)}\footnote[1]{Here we assume inductive and
coinductive data types will be distinguished by their keywords \ensuremath{\mu\cdot } and
\ensuremath{\nu\cdot }. }.

\begin{hscode}\SaveRestoreHook
\column{B}{@{}>{\hspre}l<{\hspost}@{}}%
\column{E}{@{}>{\hspre}l<{\hspost}@{}}%
\>[B]{}\mathbf{newtype}\;\mu\Sigma\mathrel{=}\Conid{In}\;\{\mskip1.5mu \Varid{in^\circ}\mathbin{::}\Sigma\;(\mu\Sigma)\mskip1.5mu\}{}\<[E]%
\ColumnHook
\end{hscode}\resethooks

A fold is a recursive function from the type \ensuremath{\mu\Sigma} to any type \ensuremath{\Varid{a}} endowed
with a function \ensuremath{\Varid{alg}\mathbin{::}\Sigma\;\Varid{a}\to \Varid{a}}. A fold takes as input the function \ensuremath{\Varid{alg}}
and recursively transforms an inductively defined data structure into an element
of type \ensuremath{\Varid{a}} using \ensuremath{\Varid{alg}} to recombine the results from the recursive calls
\begin{hscode}\SaveRestoreHook
\column{B}{@{}>{\hspre}l<{\hspost}@{}}%
\column{E}{@{}>{\hspre}l<{\hspost}@{}}%
\>[B]{}\catamor{\cdot }\mathbin{::}\Conid{Functor}\;\Sigma\Rightarrow (\Sigma\;\Varid{a}\to \Varid{a})\to \mu\Sigma\to \Varid{a}{}\<[E]%
\\
\>[B]{}\catamor{\Varid{alg}}\mathrel{=}\Varid{alg}\hsdot{\cdot}{.\ }\Varid{fmap}\;\catamor{\Varid{alg}}\hsdot{\cdot}{.\ }\Varid{in^\circ}{}\<[E]%
\ColumnHook
\end{hscode}\resethooks
For example, recall the language we introduced earlier in
Section~\ref{sec:intro}. The signature functor \ensuremath{\Sigma} of this language can be
defined as:
\begin{hscode}\SaveRestoreHook
\column{B}{@{}>{\hspre}l<{\hspost}@{}}%
\column{37}{@{}>{\hspre}l<{\hspost}@{}}%
\column{E}{@{}>{\hspre}l<{\hspost}@{}}%
\>[B]{}\mathbf{data}\;\Conid{ValAddF}\;\Varid{x}\mathrel{=}\Conid{Val}\;\Conid{Nat}\mid \Conid{Add}\;\Varid{x}\;\Varid{x}\;{}\<[37]%
\>[37]{}\mathbf{deriving}\;\Conid{Functor}{}\<[E]%
\ColumnHook
\end{hscode}\resethooks
where \ensuremath{\Conid{Val}\mathbin{::}\Conid{Nat}\to \Sigma\;\Varid{x}} and \ensuremath{\Conid{Add}\mathbin{::}\Varid{x}\to \Varid{x}\to \Sigma\;\Varid{x}} are the
constructors of the language. The language described by \ensuremath{\Sigma} is denoted by
\ensuremath{\mu\Sigma}. To simplify the presentation and avoid cluttering the code with the \ensuremath{\Conid{In}} operator we can define \emph{smart constructors} as well
\begin{hscode}\SaveRestoreHook
\column{B}{@{}>{\hspre}l<{\hspost}@{}}%
\column{52}{@{}>{\hspre}l<{\hspost}@{}}%
\column{62}{@{}>{\hspre}l<{\hspost}@{}}%
\column{E}{@{}>{\hspre}l<{\hspost}@{}}%
\>[B]{}\Varid{add}\mathbin{::}\mu\Conid{ValAddF}\to \mu\Conid{ValAddF}\to \mu\Conid{ValAddF}\;{}\<[52]%
\>[52]{}\hspace{2em}\;{}\<[62]%
\>[62]{}\Varid{val}\mathbin{::}\Nat\to \mu\Conid{ValAddF}{}\<[E]%
\\
\>[B]{}\Varid{add}\;\Varid{x}\;\Varid{y}\mathrel{=}\Conid{In}\;(\Conid{Add}\;\Varid{x}\;\Varid{y})\;{}\<[62]%
\>[62]{}\Varid{val}\mathrel{=}\Conid{In}\hsdot{\cdot}{.\ }\Conid{Val}{}\<[E]%
\ColumnHook
\end{hscode}\resethooks

The interpretation function can now be defined as a fold over a \ensuremath{\Sigma}-algebra.
For example, if we wanted to interpret this language into the domain of natural
numbers, we could use the following \ensuremath{\Sigma}-algebra on \ensuremath{\Conid{Nat}}.

\begin{hscode}\SaveRestoreHook
\column{B}{@{}>{\hspre}l<{\hspost}@{}}%
\column{17}{@{}>{\hspre}l<{\hspost}@{}}%
\column{E}{@{}>{\hspre}l<{\hspost}@{}}%
\>[B]{}\Varid{desem}\mathbin{::}\Conid{ValAddF}\;\Conid{Nat}\to \Conid{Nat}{}\<[E]%
\\
\>[B]{}\Varid{desem}\;(\Conid{Val}\;\Varid{n}){}\<[17]%
\>[17]{}\mathrel{=}\Varid{n}{}\<[E]%
\\
\>[B]{}\Varid{desem}\;(\Conid{Add}\;\Varid{n}\;\Varid{m})\mathrel{=}\Varid{n}\mathbin{+}\Varid{m}{}\<[E]%
\ColumnHook
\end{hscode}\resethooks
Then the fold over \ensuremath{\Varid{desem}}, namely \ensuremath{\catamor{\Varid{desem}}}, is by definition equal
to the interpretation function $\den{\cdot}$ we have shown in Section~\ref{sec:intro}.

\subsection{Unfolds}
\label{sec:unfolds}
While folds \emph{destruct} data, their dual, the unfolds, \emph{construct}
data.  In particular, given a seed value and an \emph{observation}, the unfold
corecursively  ``runs'' the observations starting from the initial seed.

For example, for a set of states $X$ and an alphabet set $L$ we can define a
transition system on $X$ as a function $c: X \to L \times X$ implementing the
\emph{transition} map. In particular, for a state $x_{1} \in X$,
$c(x_{1})$ returns a pair $(l, x_{2}) \in L \times X$ where $l \in L$ is the
observable action and $x_{2} \in X$ is the next state.

Now consider the transition system below composed by three states $x_{1}, x_{2}$
and $x_{3}$ with the transitions labelled by the set of labels $L = \{a, b\}$
$c_{X}$ defined such that $x_{1} \mapsto (a, x_{2})$, $x_{2} \mapsto (b, x_{3})$
and $x_{3} \mapsto (c, x_{3})$.

\[\begin{tikzcd}
    {x_1} & {x_2} & {x_3} \arrow["b"', loop, distance=2em, in=35, out=325]
	\arrow["a", from=1-1, to=1-2]
	\arrow["b", from=1-2, to=1-3]
\end{tikzcd}\]
This can be implemented by defining the \ensuremath{\Conid{Functor}} of behaviors, say \ensuremath{\Conid{BHV}}
representing the shape of the information observed in each transition and an
observation map \ensuremath{\Varid{obs}\mathbin{::}\Conid{States}\to \Conid{BHV}\;\Conid{States}} taking a starting state to the
next state together with the additional information about the observable behaviors produced in the process

\begin{hscode}\SaveRestoreHook
\column{B}{@{}>{\hspre}l<{\hspost}@{}}%
\column{12}{@{}>{\hspre}l<{\hspost}@{}}%
\column{E}{@{}>{\hspre}l<{\hspost}@{}}%
\>[B]{}\mathbf{data}\;\Conid{BHV}\;\Varid{k}\mathrel{=}\Conid{Nat}\mathbin{:<}\Varid{k}{}\<[E]%
\\
\>[B]{}\mathbf{data}\;\Conid{States}\mathrel{=}\Conid{X}_{1}\mid \Conid{X}_{2}\mid \Conid{X}_{3}{}\<[E]%
\\[\blanklineskip]%
\>[B]{}\Varid{opsem}\mathbin{::}\Conid{States}\to \Conid{BHV}\;\Conid{States}{}\<[E]%
\\
\>[B]{}\Varid{opsem}\;\Conid{X}_{1}{}\<[12]%
\>[12]{}\mathrel{=}\mathrm{1}\mathbin{:<}\Conid{X}_{2}{}\<[E]%
\\
\>[B]{}\Varid{opsem}\;\Conid{X}_{2}{}\<[12]%
\>[12]{}\mathrel{=}\mathrm{2}\mathbin{:<}\Conid{X}_{3}{}\<[E]%
\\
\>[B]{}\Varid{opsem}\;\Conid{X}_{3}{}\<[12]%
\>[12]{}\mathrel{=}\mathrm{2}\mathbin{:<}\Conid{X}_{3}{}\<[E]%
\ColumnHook
\end{hscode}\resethooks
We can set some notation to make it explicit that a this transition map is in
fact the operational semantics of our transition system
\[
  x \xrightarrow{l}_{X} x' \eqiff \ensuremath{\Varid{obs}}(x) = (l, x')
\]

In the context of transition systems, the unfold function executes the
operational semantics from a starting state while collecting the observable
behaviors that appear in each transition. Thus, we need define the type for the
collection of behaviors first.

\subsubsection{Coinductive Data Types}
Since running the operational semantics may result in non-termination we are
interested in (possibly) infinite data types. In our example above where \ensuremath{\Conid{BHV}}
was defined as the labels paired with the target state, we were interested in
the infinite collections of labels.  Hence, once the shape of the behaviours $B$
is known, we want to have the \emph{greatest} fixed-point of $B$, written
$\nu B$, and implemented as follows\footnote[1]{Notice that in Haskell,
inductive and coinductive data types coincide. Here we assume \ensuremath{\mu\cdot } and \ensuremath{\nu\cdot }
are different.}
\begin{hscode}\SaveRestoreHook
\column{B}{@{}>{\hspre}l<{\hspost}@{}}%
\column{E}{@{}>{\hspre}l<{\hspost}@{}}%
\>[B]{}\mathbf{newtype}\;\nu\Varid{b}\mathrel{=}\Conid{Out}^{\circ}\;\{\mskip1.5mu \Varid{out}\mathbin{::}\Varid{b}\;(\nu\Varid{b})\mskip1.5mu\}{}\<[E]%
\ColumnHook
\end{hscode}\resethooks
Now, for each starting element of type \ensuremath{\Varid{x}} the unfold maps this element to the unique element in \ensuremath{\nu\Varid{b}} that satisfies the following equation
\begin{hscode}\SaveRestoreHook
\column{B}{@{}>{\hspre}l<{\hspost}@{}}%
\column{E}{@{}>{\hspre}l<{\hspost}@{}}%
\>[B]{}\ana{\cdot }\mathbin{::}\Conid{Functor}\;\Varid{b}\Rightarrow (\Varid{x}\to \Varid{b}\;\Varid{x})\to \Varid{x}\to \nu\Varid{b}{}\<[E]%
\\
\>[B]{}\ana{\Varid{coalg}}\mathrel{=}\Conid{Out}^{\circ}\hsdot{\cdot}{.\ }\Varid{fmap}\;\ana{\Varid{coalg}}\hsdot{\cdot}{.\ }\Varid{coalg}{}\<[E]%
\ColumnHook
\end{hscode}\resethooks

Let us exemplify this by using the functor \ensuremath{\Conid{BHV}} of behaviors we defined above.
Its greatest fixed-point is the data type of streams (infinite lists) of numbers
\begin{hscode}\SaveRestoreHook
\column{B}{@{}>{\hspre}l<{\hspost}@{}}%
\column{E}{@{}>{\hspre}l<{\hspost}@{}}%
\>[B]{}\mathbf{type}\;\Conid{Stream}\mathrel{=}\nu\Conid{BHV}{}\<[E]%
\ColumnHook
\end{hscode}\resethooks

Now, starting from a state, say \ensuremath{\Conid{X}_{2}}, the unfold over the transition map \ensuremath{\Varid{opsem}}
is obtained by iterating \ensuremath{\Varid{opsem}} infinitely many times while collecting
observable behaviours which yields the \emph{infinite trace}
``$\ensuremath{\mathrm{22222222222222}}\dots$'' for the state \ensuremath{\Conid{X}_{2}}.

\subsubsection{Full Abstraction for unfolds}
One can easily see that despite \ensuremath{\Conid{X}_{2}} and \ensuremath{\Conid{X}_{3}} are different states, their trace
is equal. That is \ensuremath{\ana{\Varid{opsem}}\;\Conid{X}_{2}=\ana{\Varid{opsem}}\;\Conid{X}_{3}}. This means that
that two states produce the same trace if and only if they are mapped into the
same streams. To see this, consider the state \ensuremath{\Conid{X}_{2}} again and notice that \ensuremath{\Varid{opsem}}
takes \ensuremath{\Conid{X}_{2}} to \ensuremath{\Conid{X}_{3}} with label \ensuremath{\mathrm{2}} then \ensuremath{\ana{\Varid{opsem}}\;\Conid{X}_{2}} must map \ensuremath{\Conid{X}_{2}} to an
infinite stream such that the first label is \ensuremath{\mathrm{2}} and the rest of the stream is
where \ensuremath{\Conid{X}_{3}} is mapped to, hence the unfold preserves the behaviors when mapping
into \ensuremath{\nu\Varid{b}}, so if two states are producing the same trace (operationally) so
must be their unfolding. The other direction is provided by the fact that the
unfold maps every state to its \emph{unique} trace. In other words, there is no
other trace that can represent one state so it the traces of two states are
equal then their states behave the same operationally. For a more formal
explanation of this phenomenon, we defer the interested reader to
~\cite[Appendix, Theorem 1]{Hinze08}.

\subsubsection{Other behavioral functors}
Depending on the functor $B$ we choose we can model different kind of
semantics.

Another example is big-step operational semantics which is modelled by the
constant functor \ensuremath{\Conid{Big}\;\Varid{x}\mathrel{=}\Conid{V}} where \ensuremath{\Conid{V}} is the set of values for a language. Its
associated \ensuremath{\Conid{V}}-coalgebra \ensuremath{\cdot \Downarrow\cdot \mathbin{::}\mu\Sigma\to \Conid{V}} sends programs in to the
values in $V$ they reduce to. We can use classic notation for big-step semantics
as follows
\[
  t \bigstep v \eqiff \ensuremath{\cdot \Downarrow\cdot }(t) = v
\]

So far we have considered semantics for deterministic languages.  However, when
the language is non-deterministic we need a way to map a program into multiple
programs. To this end we can use the finite powerset functor $\finpow{-}$ which
can be implemented in functional programming using the type of lists.  In this
case, the operational semantics is a function of type \ensuremath{\Varid{x}\to [\mskip1.5mu \Varid{x}\mskip1.5mu]} or even \ensuremath{\Varid{x}\to [\mskip1.5mu (\Conid{L},\Varid{x})\mskip1.5mu]} where each transition is annotated with a label in the set \ensuremath{\Conid{L}}.

As in the previous cases, we can set some notation as below stating that $x_{1}$
reduces to $x_{2}$ if there exists $x_{2}$ in the list of states $x_{1}$ maps to.
\[
 x_{1}  \to_{X} x_{2} \eqiff  x_{2} \in \ensuremath{\Varid{opsem}}(x_{1})
\]

Now the we introduced both folds and unfolds and how to use unfolds on
transition systems we need to put these two pieces together and this is the
topic of the next section.

\newcommand{\stopval}[1]{\underline{#1}}

\section{Simple Recursion Schemes for Program Semantics}
\label{sec:simple-laws}\label{sec:simple-sos}
In the previous section we have shown how to give a compositional denotational
semantics for a small programming language by using folds
(Section~\ref{sec:folds}), how coalgebras can be used to describe operational
semantics and, furthermore, how unfolds are fully abstract interpretations of
transition systems (Section~\ref{sec:unfolds}).

Our aim is now to get compositionality and full abstraction with the same
interpretation function. We do this by using a distributive law between syntax
and semantics. In this section we first exemplify distributive laws via the
arithmetic language (Section~\ref{sec:simple-distr-laws}) and then we generalise
the framework to a generic signature and behavioral functor
(Section~\ref{sec:simple-distr-laws-gen}).

\subsection{Distributive Laws}
\label{sec:simple-distr-laws}
To understand how this is useful to achieve our goal consider the signature and
the behavioral functor we defined in Section~\ref{sec:background}, in
particular, \ensuremath{\Conid{ValAddF}} and \ensuremath{\Conid{BHV}}.

Say that we want to define an operational semantics for the language described
by \ensuremath{\mu\Conid{ValAddF}} and, in particular, we would like to have a semantics that
behaves as per the following inference rules:
\begin{mathpar}
  \inferrule{ }{\ensuremath{\Varid{val}\;\Varid{n}} \reduceto{n} \ensuremath{\Varid{val}\;\Varid{n}}}
  \and
  \inferrule{t1 \reduceto{n} t1' \qquad \ensuremath{\Varid{t}_{2}}  \reduceto{m} \ensuremath{\Varid{t}_{2}'}}{\ensuremath{\Varid{add}\;\Varid{t}_{1}\;\Varid{t}_{2}} \reduceto{n + m} \ensuremath{\Varid{add}\;\Varid{t}_{1}'\;\Varid{t}_{2}'}}
\end{mathpar}
Mathematically, $\xrightarrow{n}$ is a inductively defined family of relations
which happens to be a function as well. In particular, it is defined inductively
over the structure of the language mapping a value \ensuremath{\Varid{val}\;\Varid{n}} to \ensuremath{(\Varid{n},\Varid{val}\;\Varid{n})} and
a program \ensuremath{\Varid{add}\;\Varid{t}_{1}\;\Varid{t}_{2}} to \ensuremath{(\Varid{n}\mathbin{+}\Varid{m},\Varid{add}\;\Varid{t}_{1}'\;\Varid{t}_{2}')} if the two subprograms \ensuremath{\Varid{t}_{1}} and
\ensuremath{\Varid{t}_{2}} map to \ensuremath{(\Varid{n},\Varid{t}_{1}')} and \ensuremath{(\Varid{m},\Varid{t}_{2}')} respectively.

It is easy to see that the above specification corresponds to an observation
function over \ensuremath{\mu\Conid{ValAddF}}, i.e. a function of type \ensuremath{\mu\Conid{ValAddF}\to \Conid{BHV}\;(\mu\Conid{ValAddF})} sending a program to another program together with an observation as
in the following recursive program:
\begin{hscode}\SaveRestoreHook
\column{B}{@{}>{\hspre}l<{\hspost}@{}}%
\column{33}{@{}>{\hspre}l<{\hspost}@{}}%
\column{35}{@{}>{\hspre}l<{\hspost}@{}}%
\column{39}{@{}>{\hspre}l<{\hspost}@{}}%
\column{E}{@{}>{\hspre}l<{\hspost}@{}}%
\>[B]{}\Varid{opsemSimple}\mathbin{::}\mu\Conid{ValAddF}\to \Conid{BHV}\;(\mu\Conid{ValAddF}){}\<[E]%
\\
\>[B]{}\Varid{opsemSimple}\;(\Conid{In}\;(\Conid{Val}\;\Varid{n})){}\<[33]%
\>[33]{}\mathrel{=}\Varid{n}\mathbin{:<}\Conid{In}\;(\Conid{Val}\;\Varid{n}){}\<[E]%
\\
\>[B]{}\Varid{opsemSimple}\;(\Conid{In}\;(\Conid{Add}\;\Varid{t}_{1}\;\Varid{t}_{2})){}\<[33]%
\>[33]{}\mathrel{=}\mathbf{let}\;(\Varid{n}\mathbin{:<}\Varid{t}_{1}')\mathrel{=}\Varid{opsemSimple}\;\Varid{t}_{1}{}\<[E]%
\\
\>[33]{}\hsindent{6}{}\<[39]%
\>[39]{}(\Varid{m}\mathbin{:<}\Varid{t}_{2}')\mathrel{=}\Varid{opsemSimple}\;\Varid{t}_{2}{}\<[E]%
\\
\>[33]{}\hsindent{2}{}\<[35]%
\>[35]{}\mathbf{in}{}\<[E]%
\\
\>[35]{}\hsindent{4}{}\<[39]%
\>[39]{}(\Varid{n}\mathbin{+}\Varid{m})\mathbin{:<}\Varid{add}\;\Varid{t}_{1}'\;\Varid{t}_{2}'{}\<[E]%
\ColumnHook
\end{hscode}\resethooks
Since this is a recursive definition where the recursive call is only used on a
strictly smaller input we can equivalently write \ensuremath{\Varid{opsem}} as a fold over an
\ensuremath{\Conid{ValAddF}}-algebra on \ensuremath{\Conid{BHV}\;(\mu\Conid{ValAddF})}, i.e. a map \ensuremath{\Conid{ValAddF}\;(\Conid{BHV}\;(\mu\Conid{ValAddF}))\to \Conid{BHV}\;(\mu\Conid{ValAddF})} as in the program below:
\begin{hscode}\SaveRestoreHook
\column{B}{@{}>{\hspre}l<{\hspost}@{}}%
\column{38}{@{}>{\hspre}l<{\hspost}@{}}%
\column{E}{@{}>{\hspre}l<{\hspost}@{}}%
\>[B]{}\Varid{opsemAlg}\mathbin{::}\Conid{ValAddF}\;(\Conid{BHV}\;(\mu\Conid{ValAddF}))\to \Conid{BHV}\;(\mu\Conid{ValAddF}){}\<[E]%
\\
\>[B]{}\Varid{opsemAlg}\;(\Conid{Val}\;\Varid{n}){}\<[38]%
\>[38]{}\mathrel{=}\Varid{n}\mathbin{:<}\Conid{In}\;(\Conid{Val}\;\Varid{n}){}\<[E]%
\\
\>[B]{}\Varid{opsemAlg}\;(\Conid{Add}\;(\Varid{n}\mathbin{:<}\Varid{t}_{1}')\;(\Varid{m}\mathbin{:<}\Varid{t}_{2}'))\mathrel{=}(\Varid{n}\mathbin{+}\Varid{m})\mathbin{:<}\Conid{In}\;(\Conid{Add}\;\Varid{t}_{1}'\;\Varid{t}_{2}'){}\<[E]%
\ColumnHook
\end{hscode}\resethooks
Next we notice that this algebra can be further decomposed by noticing that \ensuremath{\Conid{In}}
is applied to the target program of type \ensuremath{\Conid{ValAddF}\;(\mu\Conid{ValAddF})}. In fact we can
remove \ensuremath{\Conid{In}} resulting in an alternative definition of \ensuremath{\Varid{opsemAlg}} which we call
\ensuremath{\Varid{opsemDistr}}
\begin{hscode}\SaveRestoreHook
\column{B}{@{}>{\hspre}l<{\hspost}@{}}%
\column{40}{@{}>{\hspre}l<{\hspost}@{}}%
\column{E}{@{}>{\hspre}l<{\hspost}@{}}%
\>[B]{}\Varid{opsemDistr}\mathbin{::}\Conid{ValAddF}\;(\Conid{BHV}\;(\mu\Conid{ValAddF}))\to \Conid{BHV}\;(\Conid{ValAddF}\;(\mu\Conid{ValAddF})){}\<[E]%
\\
\>[B]{}\Varid{opsemDistr}\;(\Conid{Val}\;\Varid{n}){}\<[40]%
\>[40]{}\mathrel{=}\Varid{n}\mathbin{:<}\Conid{Val}\;\Varid{n}{}\<[E]%
\\
\>[B]{}\Varid{opsemDistr}\;(\Conid{Add}\;(\Varid{n}\mathbin{:<}\Varid{t}_{1}')\;(\Varid{m}\mathbin{:<}\Varid{t}_{2}'))\mathrel{=}(\Varid{n}\mathbin{+}\Varid{m})\mathbin{:<}\Conid{Add}\;\Varid{t}_{1}'\;\Varid{t}_{2}'{}\<[E]%
\ColumnHook
\end{hscode}\resethooks
which post-composed with \ensuremath{\Varid{fmap}\;\Conid{In}} will give \ensuremath{\Varid{opsemAlg}}. It is an easy
simplification now to parametrise \ensuremath{\Varid{opsemDistr}} by substituting \ensuremath{\mu\Conid{ValAddF}}
with a universally quantified type variable, thus  obtaining the following
parametric function:
\begin{hscode}\SaveRestoreHook
\column{B}{@{}>{\hspre}l<{\hspost}@{}}%
\column{E}{@{}>{\hspre}l<{\hspost}@{}}%
\>[B]{}\Varid{opsemDistr}\mathbin{::}\forall \Varid{x}\hsforall \hsdot{\cdot}{.\ }\Conid{ValAddF}\;(\Conid{BHV}\;\Varid{x})\to \Conid{BHV}\;(\Conid{ValAddF}\;\Varid{x}){}\<[E]%
\ColumnHook
\end{hscode}\resethooks
Notice how this does not change the implementation of \ensuremath{\Varid{opsemDistr}} since when we
defined \ensuremath{\Varid{opsemDistr}} we did no rely on the structure of \ensuremath{\mu\Conid{ValAddF}}.

The above study case dualises nicely, of course. In fact, the denotational
semantics is an \ensuremath{\Conid{ValAddF}}-algebra over the type \ensuremath{\nu\Conid{B}} defined by \ensuremath{\Varid{dsem}\mathrel{=}\Varid{opsemDistr}\hsdot{\cdot}{.\ }\Varid{fmap}\;\Varid{out}}.

\subsection{The general case}
\label{sec:simple-distr-laws-gen}
In the example above, \ensuremath{\Varid{opsemDistr}} is a distributive law between the signature functor
\ensuremath{\Conid{ValAddF}} and the behavioural functor \ensuremath{\Conid{BHV}} .

In general, given a signature and a behavior functor $\Sigma$ and \ensuremath{\Varid{b}} respectively,
a \emph{distributive law} is a parametric function distributing \ensuremath{\Sigma} over \ensuremath{\Varid{b}} as follows:
\begin{hscode}\SaveRestoreHook
\column{B}{@{}>{\hspre}l<{\hspost}@{}}%
\column{E}{@{}>{\hspre}l<{\hspost}@{}}%
\>[B]{}\lambda\mathbin{::}(\Conid{Functor}\;\Sigma,\Conid{Functor}\;\Varid{b})\Rightarrow \forall \Varid{x}\hsforall \hsdot{\cdot}{.\ }\Sigma\;(\Varid{b}\;\Varid{x})\to \Varid{b}\;(\Sigma\;\Varid{x}){}\<[E]%
\ColumnHook
\end{hscode}\resethooks
At this point, given a distributive law \ensuremath{\lambda} between \ensuremath{\Sigma} and \ensuremath{\Varid{b}}, the following facts hold
\begin{itemize}
  \item the operational model is a \ensuremath{\Varid{b}}-coalgebra of type \ensuremath{\mu\Sigma\to \Varid{b}\;(\mu\Sigma)} given by a fold over the  \ensuremath{\Sigma}-algebra \ensuremath{(\Varid{fmap}\;\Conid{In}\hsdot{\cdot}{.\ }\lambda)\mathbin{::}\Sigma\;(\Varid{b}\;(\mu\Sigma))\to \Varid{b}\;(\mu\Sigma)}
\begin{hscode}\SaveRestoreHook
\column{B}{@{}>{\hspre}l<{\hspost}@{}}%
\column{E}{@{}>{\hspre}l<{\hspost}@{}}%
\>[B]{}\Varid{opsem}\;\lambda\mathrel{=}\catamor{\Varid{fmap}\;\Conid{In}\hsdot{\cdot}{.\ }\lambda}{}\<[E]%
\ColumnHook
\end{hscode}\resethooks
  \item dually, the denotational model is a \ensuremath{\Sigma}-algebra of type \ensuremath{\Sigma\;(\nu\Varid{b})\to \nu\Varid{b}} given by the unfold over the \ensuremath{\Varid{b}}-coalgebra \ensuremath{(\lambda\hsdot{\cdot}{.\ }\Varid{fmap}\;\Varid{out})\mathbin{::}\Sigma\;(\nu\Varid{b})\to \Varid{b}\;(\Sigma\;(\nu\Varid{b}))}
\begin{hscode}\SaveRestoreHook
\column{B}{@{}>{\hspre}l<{\hspost}@{}}%
\column{E}{@{}>{\hspre}l<{\hspost}@{}}%
\>[B]{}\Varid{desem}\;\lambda\mathrel{=}\ana{\lambda\hsdot{\cdot}{.\ }\Varid{fmap}\;\Varid{out}}{}\<[E]%
\ColumnHook
\end{hscode}\resethooks
  \item the fold over the denotational model is compositional
  \item unfold over the operational model is fully abstract      
\end{itemize}
Furthermore, the fold over the denotational model corresponds to the unfold over
the operational model. This map is called the \emph{universal
semantics}~(\cite{TuriP97}).
\begin{hscode}\SaveRestoreHook
\column{B}{@{}>{\hspre}l<{\hspost}@{}}%
\column{46}{@{}>{\hspre}l<{\hspost}@{}}%
\column{E}{@{}>{\hspre}l<{\hspost}@{}}%
\>[B]{}\Varid{sem}\mathbin{::}\forall \Sigma\hsforall \;\Varid{b}\hsdot{\cdot}{.\ }(\Conid{Functor}\;\Sigma,\Conid{Functor}\;\Varid{b})\Rightarrow (\forall \Varid{x}\hsforall \hsdot{\cdot}{.\ }\Sigma\;(\Varid{b}\;\Varid{x})\to \Varid{b}\;(\Sigma\;\Varid{x}))\to \mu\Sigma\to \nu\Varid{b}{}\<[E]%
\\
\>[B]{}\Varid{sem}\;\lambda\;\Varid{t}\mathrel{=}\catamor{\Varid{desem}\;\lambda}\;\Varid{t}={}\<[46]%
\>[46]{}\ana{\Varid{opsem}\;\lambda}\;\Varid{t}{}\<[E]%
\ColumnHook
\end{hscode}\resethooks
As shorthand, we use semantic brackets for this definition: \ensuremath{\den{\Varid{t}}{\lambda}\mathrel{=}\Varid{sem}\;\lambda\;\Varid{t}}.

As a last remark of this section, it is worthwhile to notice the correspondence
between distributive laws and certain \emph{rule formats}. The kind of
distributive laws we have seen in section, in particular, correspond to
Structural Operational Semantics (SOS) (\cite{Plotkin04a}). For example, for the
behavioral functor \ensuremath{\Conid{BHV}} and a generic signature functor the distributive law
yields an operational semantics of the following shape:
\begin{equation}
  \label{rule:simple}
  \inferrule{x_{1}\reduceto{l_{1}} x'_{1} \quad  \dots \quad x_{n} \reduceto{l_{n}} x'_{n}}{\sigma(x_1, \cdots, x_n) \reduceto{l} \sigma'(y_1, \cdots, y_m)}
\end{equation}
for a set of meta-variables
$\{y_{1} \dots, y_{m} \} \subseteq \{x'_{1}, \dots, x'_{n}\}$, constructors from
the signature $\sigma, \sigma' \in \Sigma$ and labels
$\{l_1, \dots, l_{n}, l\} \in \ensuremath{\Conid{L}} $. Notice that the distributive law does not
force the labels in the conclusions to belong to those in the premises, but
there are however restrictions on the variables in the target belonging to the
targets in the premises. We will see in the next sections that these restriction
limit the expressivity of the languages we can model.

\section{Full Abstraction for a Simple Non-Deterministic Arithmetic Language}
\label{sec:copointed}
The language we introduced in \autoref{sec:simple-sos} is a basic language for
talking about streams. We would like now to implement a fully abstract
interpretation for the language we have introduced in \autoref{sec:intro}, which
is the language of arithmetic expressions originally introduced by
\citet{Hutton98}, also known as Hutton's razor.

In \autoref{sec:razor} we introduce Hutton's razor while in
\autoref{sec:razor-full-abstrac} we show how to give a fully abstract semantics
and in \autoref{sec:copointed-functors} we abstract to copointed functors to
define the behavior of these kind of languages. The full general case here is
deferred to \autoref{sec:ccs}.

\subsection{Hutton's razor}
\label{sec:razor}
This language has the same syntax we have seen in \autoref{sec:simple-sos} so we
can just reuse the signature functor \ensuremath{\Conid{ValAddF}}.
The operational semantics of Hutton's razor is defined using the following
inductive relation $({\to}) \subseteq \mu \Sigma \times \mu \Sigma$:
\begin{mathpar}
  \inferrule{ }{\operator{add}(\operator{val}(n),\operator{val}(m)) \reduceto{} \operator{val}(n + m)} \and
  \inferrule{t_{1} \reduceto{} t'_{1}}{\operator{add}(t_{1},t_{2}) \reduceto{} \operator{add}(t'_{1},t_{2})} \and
  \inferrule{t_{2} \reduceto{} t'_{2}}{\operator{add}(t_{1},t_{2}) \reduceto{} \operator{add}(t_{1},t'_{2})}
\end{mathpar}
The semantics of this language is clearly non-deterministic as the operation
$\operator{add}$ can perform two different reductions when the inner term is not
$\operator{val}$.  Thus, to model it, we use an observation function for the
functor $\finpow{-}$ which we implement using finite lists of programs. These
lists should be compared modulo duplication and swapping of elements.  With this
in mind the implementation of the inference rules above is the following:
\begin{hscode}\SaveRestoreHook
\column{B}{@{}>{\hspre}l<{\hspost}@{}}%
\column{49}{@{}>{\hspre}l<{\hspost}@{}}%
\column{51}{@{}>{\hspre}l<{\hspost}@{}}%
\column{E}{@{}>{\hspre}l<{\hspost}@{}}%
\>[B]{}\Varid{smallstep}\mathbin{::}\mu\Conid{ValAddF}\to [\mskip1.5mu \mu\Conid{ValAddF}\mskip1.5mu]{}\<[E]%
\\
\>[B]{}\Varid{smallstep}\;(\Conid{In}\;(\Conid{Val}\;\Varid{n})){}\<[49]%
\>[49]{}\mathrel{=}[\mskip1.5mu \mskip1.5mu]{}\<[E]%
\\
\>[B]{}\Varid{smallstep}\;(\Conid{In}\;(\Conid{Add}\;(\Conid{In}\;(\Conid{Val}\;\Varid{n}))\;(\Conid{In}\;(\Conid{Val}\;\Varid{m})))){}\<[49]%
\>[49]{}\mathrel{=}[\mskip1.5mu (\Varid{val}\;(\Varid{n}\mathbin{+}\Varid{m}))\mskip1.5mu]{}\<[E]%
\\
\>[B]{}\Varid{smallstep}\;(\Conid{In}\;(\Conid{Add}\;\Varid{t}_{1}\;\Varid{t}_{2})){}\<[49]%
\>[49]{}\mathrel{=}[\mskip1.5mu \Varid{add}\;\Varid{t}_{1}'\;\Varid{t}_{2}\mid \Varid{t}_{1}'\leftarrow \Varid{smallstep}\;\Varid{t}_{1}\mskip1.5mu]{}\<[E]%
\\
\>[49]{}\hsindent{2}{}\<[51]%
\>[51]{}\mathbin{+\!\!+}[\mskip1.5mu \Varid{add}\;\Varid{t}_{1}\;\Varid{t}_{2}'\mid \Varid{t}_{2}'\leftarrow \Varid{smallstep}\;\Varid{t}_{2}\mskip1.5mu]{}\<[E]%
\ColumnHook
\end{hscode}\resethooks
Notice how the base case for \ensuremath{\Varid{val}} is empty as there is no reduction of the \ensuremath{\Varid{val}} case.

There are essentially two problems with this implementation. The first is that
the unfold over this observation function yields the finitely branching
\emph{empty} trees. It is an easy computation to perform by unfolding the
definition of, for example, \ensuremath{\ana{\Varid{smallstep}}\;(\Varid{add}\;(\Varid{val}\;\Varid{n})\;(\Varid{val}\;\Varid{m}))}.
Intuitively, this because after one step of computation we end up with the value
\ensuremath{[\mskip1.5mu \Varid{val}\;(\Varid{n}\mathbin{+}\Varid{m})\mskip1.5mu]} by definition of the second case and if we perform another step of
computation by applying \ensuremath{\ana{\Varid{smallstep}}} on the programs of the list, by the
first case these (in this case we have just one program) will reduce to the
empty list.

The second problem is that, in order to reap the benefits of Turi and Plotkin's
approach we would need a parametric distributive law. This is rather problematic
because, in the implementation of \ensuremath{\lambda} below, it is not clear what should go
in the holes ``\ensuremath{\anonymous }'':
\begin{hscode}\SaveRestoreHook
\column{B}{@{}>{\hspre}l<{\hspost}@{}}%
\column{70}{@{}>{\hspre}l<{\hspost}@{}}%
\column{E}{@{}>{\hspre}l<{\hspost}@{}}%
\>[B]{}\lambda\mathbin{::}\Conid{ValAddF}\;([\mskip1.5mu \Varid{x}\mskip1.5mu])\to [\mskip1.5mu (\Conid{ValAddF}\;\Varid{x})\mskip1.5mu]{}\<[E]%
\\
\>[B]{}\lambda\;(\Conid{Add}\;(\Varid{n}\mathbin{:<}\Varid{xs})\;(\Varid{m}\mathbin{:<}\Varid{ys}))\mathrel{=}([\mskip1.5mu \Conid{Add}\;\Varid{x}\;\anonymous \mid \Varid{x}\leftarrow \Varid{xs}\mskip1.5mu]\mathbin{+\!\!+}[\mskip1.5mu \Conid{Add}\;\anonymous \;\Varid{y}{}\<[70]%
\>[70]{}\mid \Varid{y}\leftarrow \Varid{ys}\mskip1.5mu]){}\<[E]%
\ColumnHook
\end{hscode}\resethooks
In fact, in the case of \ensuremath{\Conid{Add}\;(\Varid{n}\mathbin{:<}\Varid{xs})\;(\Varid{m}\mathbin{:<}\Varid{ys})} the two subprograms have already
generated their list of outputs of type \ensuremath{\Varid{x}} and since \ensuremath{\Varid{x}} is universally quantified
we do not know which element of \ensuremath{\Varid{x}} actually generated this reduction which
is what we need to complete the holes.

\subsection{Full abstraction for the razor}
\label{sec:razor-full-abstrac}
The first problem can be addressed by tweaking the definition of the behavioral
functor. The idea is to make explicit when the semantics is proceeding and when
it stops. In particular, we would like to map \ensuremath{\Varid{val}\;\Varid{n}} to the number \ensuremath{\Varid{n}} so that
the reduction semantics for values is not empty.

Thus, the first tweak is to use the behavioral functor $\Nat + \finpow{-}$ where
the left injection into the sum is the when the semantics stops with a number
and the right injection is when the semantics is reducing
(non-deterministically) to a program. This can be implemented as follows:
\begin{hscode}\SaveRestoreHook
\column{B}{@{}>{\hspre}l<{\hspost}@{}}%
\column{E}{@{}>{\hspre}l<{\hspost}@{}}%
\>[B]{}\mathbf{data}\;\Conid{StopAndGo}\;\Varid{k}\mathrel{=}\underline{\Conid{Nat}}\mid \Conid{Step}\;[\mskip1.5mu \Varid{k}\mskip1.5mu]\;\mathbf{deriving}\;\Conid{Functor}{}\<[E]%
\ColumnHook
\end{hscode}\resethooks
Notice here that the underscore is the constructor of the data type.

The second tweak is more is more profound as we need to modify the type of the
distributive law. In particular, we would like to retain a copy of the input
before it gets destructured. To do this, we pair the type \ensuremath{\Conid{StopAndGo}\;\Varid{x}} with an
additional \ensuremath{\Varid{x}} which is supposed to retain the original copy of the input.
Notice that the resulting type \ensuremath{(\Varid{x},\Conid{StopAndGo}\;\Varid{x})} is also a \ensuremath{\Conid{Functor}} on \ensuremath{\Varid{x}}.

\begin{hscode}\SaveRestoreHook
\column{B}{@{}>{\hspre}l<{\hspost}@{}}%
\column{E}{@{}>{\hspre}l<{\hspost}@{}}%
\>[B]{}\rho\mathbin{::}\forall \Varid{x}\hsforall \hsdot{\cdot}{.\ }\Conid{ValAddF}\;(\Varid{x},\Conid{StopAndGo}\;\Varid{x})\to \Conid{StopAndGo}\;(\Conid{ValAddF}\;\Varid{x}){}\<[E]%
\ColumnHook
\end{hscode}\resethooks
This new structure we have just inserted ensures that when the $\rho$-rule needs
to repeat the input as is, it just needs to use the left projection.  

However, it has to be noted that this is not a distributive law anymore since
\ensuremath{(\Varid{x},\Conid{StopAndGo}\;\Varid{x})} and \ensuremath{\Conid{StopAndGo}\;\Varid{x}} are not the same. To avoid confusion, from
here on we call this type of parametric function a $\rho$-rule (following
original work by Turi and Plotkin). Notice also that this $\rho$-rule not being
a distributive law has some theoretical consequences that we are going to need
to address. For the time being let us look at how we can quickly implement the
operational semantics we are after. So consider first the following
implementation of a $\rho$-rule
\begin{hscode}\SaveRestoreHook
\column{B}{@{}>{\hspre}l<{\hspost}@{}}%
\column{28}{@{}>{\hspre}l<{\hspost}@{}}%
\column{47}{@{}>{\hspre}l<{\hspost}@{}}%
\column{55}{@{}>{\hspre}l<{\hspost}@{}}%
\column{E}{@{}>{\hspre}l<{\hspost}@{}}%
\>[B]{}\rho\;(\Conid{Val}\;\Varid{n}){}\<[47]%
\>[47]{}\mathrel{=}\underline{\Varid{n}}{}\<[E]%
\\
\>[B]{}\rho\;(\Conid{Add}\;(\anonymous ,\underline{\Varid{n}})\;{}\<[28]%
\>[28]{}(\anonymous ,\underline{\Varid{m}})){}\<[47]%
\>[47]{}\mathrel{=}\underline{\Varid{n}\mathbin{+}\Varid{m}}{}\<[E]%
\\
\>[B]{}\rho\;(\Conid{Add}\;(\Varid{x}_{1},(\Conid{Step}\;\Varid{xs}_{1}))\;{}\<[28]%
\>[28]{}(\Varid{x}_{2},(\Conid{Step}\;\Varid{xs}_{2}))){}\<[47]%
\>[47]{}\mathrel{=}\Conid{Step}\;([\mskip1.5mu \Conid{Add}\;\Varid{x}_{1}\;\Varid{x}_{2}'\mid \Varid{x}_{2}'\leftarrow \Varid{xs}_{2}\mskip1.5mu]\mathbin{+\!\!+}{}\<[E]%
\\
\>[47]{}\hsindent{8}{}\<[55]%
\>[55]{}[\mskip1.5mu \Conid{Add}\;\Varid{x}_{1}'\;\Varid{x}_{2}\mid \Varid{x}_{1}'\leftarrow \Varid{xs}_{1}\mskip1.5mu]){}\<[E]%
\\
\>[B]{}\rho\;(\Conid{Add}\;(\Varid{x}_{1},\anonymous )\;(\anonymous ,(\Conid{Step}\;\Varid{xs}_{2}))){}\<[47]%
\>[47]{}\mathrel{=}\Conid{Step}\;[\mskip1.5mu \Conid{Add}\;\Varid{x}_{1}\;\Varid{x}_{2}'\mid \Varid{x}_{2}'\leftarrow \Varid{xs}_{2}\mskip1.5mu]{}\<[E]%
\\
\>[B]{}\rho\;(\Conid{Add}\;(\anonymous ,(\Conid{Step}\;\Varid{xs}_{1}))\;(\Varid{x}_{2},\anonymous )){}\<[47]%
\>[47]{}\mathrel{=}\Conid{Step}\;[\mskip1.5mu \Conid{Add}\;\Varid{x}_{1}'\;\Varid{x}_{2}\mid \Varid{x}_{1}'\leftarrow \Varid{xs}_{1}\mskip1.5mu]{}\<[E]%
\ColumnHook
\end{hscode}\resethooks
Note how this additional argument plays the role of copy of the input, for
example in the third line of this definition where \ensuremath{\Varid{x}_{1}} is the original program
that can reduced to a list of programs \ensuremath{\Varid{xs}_{1}} and in this case, since we are not
interested in this information we may retain the original program.
The relational definition corresponding to this $\rho$-rule is the following:
\begin{mathpar}
  \inferrule{ }{\ensuremath{\Varid{val}\;\Varid{n}} \reduceto{} \ensuremath{\underline{\Varid{n}}}}
  \quad
  \inferrule{\ensuremath{\Varid{t}_{1}} \reduceto{} \ensuremath{\underline{\Varid{n}}} \qquad \ensuremath{\Varid{t}_{2}} \reduceto{} \ensuremath{\underline{\Varid{m}}}}{\ensuremath{\Varid{add}\;\Varid{t}_{1}\;\Varid{t}_{2}} \reduceto{} \ensuremath{\underline{\Varid{n}\mathbin{+}\Varid{m}}}}
  \quad
  \inferrule{\ensuremath{\Varid{t}_{1}} \reduceto{} \ensuremath{\Varid{t}_{1}'}}{\ensuremath{\Varid{add}\;\Varid{t}_{1}\;\Varid{t}_{2}} \reduceto{} \ensuremath{\Varid{add}\;\Varid{t}_{1}'\;\Varid{t}_{2}}}
  \quad
  \inferrule{\ensuremath{\Varid{t}_{2}} \reduceto{} \ensuremath{\Varid{t}_{2}'}}{\ensuremath{\Varid{add}\;\Varid{t}_{1}\;\Varid{t}_{2}} \reduceto{} \ensuremath{\Varid{add}\;\Varid{t}_{1}\;\Varid{t}_{2}'}}
\end{mathpar}

It is probably worth noticing how we started with a set of rules, the arithmetic expression language by Hutton, and using distributive laws forced us to consider a slight different version of the operational semantics. This kind of format corresponds to the copointed structural operational semantics (Copointed SOS) due to the use of the copointed functor in the distributive law.

\subsection{On CoPointed Functors}
\label{sec:copointed-functors}
The type \ensuremath{(\Varid{x},\Varid{b}\;\Varid{x})} is a \ensuremath{\Conid{Functor}} on \ensuremath{\Varid{x}} and, in particular, is the \emph{free
copointed functor} for $b$. The terminology is borrowed from its dual, the
\emph{free pointed functor} \ensuremath{\Conid{Either}\;\Varid{x}\;(\Varid{b}\;\Varid{x})} since $b$ is augmented with an
extra point.
\begin{hscode}\SaveRestoreHook
\column{B}{@{}>{\hspre}l<{\hspost}@{}}%
\column{E}{@{}>{\hspre}l<{\hspost}@{}}%
\>[B]{}\mathbf{type}\;\copointed{\Varid{b}}\;\Varid{x}\mathrel{=}(\Varid{x},\Varid{b}\;\Varid{x}){}\<[E]%
\ColumnHook
\end{hscode}\resethooks
The type \ensuremath{\copointed{\cdot }} is clearly a functor with the obvious \ensuremath{\Varid{pmap}} function:
\begin{hscode}\SaveRestoreHook
\column{B}{@{}>{\hspre}l<{\hspost}@{}}%
\column{9}{@{}>{\hspre}l<{\hspost}@{}}%
\column{E}{@{}>{\hspre}l<{\hspost}@{}}%
\>[B]{}\Varid{pmap}\mathbin{::}\Conid{Functor}\;\Varid{b}\Rightarrow (\Varid{x}\to \Varid{y})\to \copointed{\Varid{b}}\;\Varid{x}\to \copointed{\Varid{b}}\;\Varid{y}{}\<[E]%
\\
\>[B]{}\Varid{pmap}\;\Varid{f}\;{}\<[9]%
\>[9]{}(\Varid{x},\Varid{y})\mathrel{=}(\Varid{f}\;\Varid{x},\Varid{fmap}\;\Varid{f}\;\Varid{y}){}\<[E]%
\ColumnHook
\end{hscode}\resethooks
There is also an obvious counit map which is simply returning the copy of the
input:
\begin{hscode}\SaveRestoreHook
\column{B}{@{}>{\hspre}l<{\hspost}@{}}%
\column{E}{@{}>{\hspre}l<{\hspost}@{}}%
\>[B]{}\epsilon\mathbin{::}\copointed{\Varid{b}}\;\Varid{a}\to \Varid{a}{}\<[E]%
\\
\>[B]{}\epsilon\;(\Varid{x},\anonymous )\mathrel{=}\Varid{x}{}\<[E]%
\ColumnHook
\end{hscode}\resethooks

Given an observation function for the functor \ensuremath{\Varid{b}}, say \ensuremath{\Varid{g}\mathbin{::}\Varid{a}\to \Varid{b}\;\Varid{a}}, an \emph{observation function for the copointed functor} \ensuremath{\copointed{\Varid{b}}} is a function of type \ensuremath{\Varid{id}\wedge\Varid{g}\mathbin{::}\Varid{a}\to \copointed{\Varid{b}}\;\Varid{a}} which pairs up \ensuremath{\Varid{g}} with the identity on \ensuremath{\Varid{a}}. In other words, observations for
copointed functors and their respective functors are in one-to-one
correspondence\footnote[1]{This is a consequence of the fact that the category
$B$-Alg is isomorphic to the category $\Id \times B$-Alg of algebras for
copointed functors}. This is implemented via \ensuremath{\floorCoPointed{\cdot }} and
\ensuremath{\ceilingCoPointed{\cdot }}
\begin{hscode}\SaveRestoreHook
\column{B}{@{}>{\hspre}l<{\hspost}@{}}%
\column{E}{@{}>{\hspre}l<{\hspost}@{}}%
\>[B]{}\floorCoPointed{\cdot }\mathbin{::}(\Conid{Functor}\;\Varid{g})\Rightarrow (\Varid{a}\to \copointed{\Varid{g}}\;\Varid{a})\to \Varid{a}\to \Varid{g}\;\Varid{a}{}\<[E]%
\\
\>[B]{}\floorCoPointed{\Varid{ccoalg}}\;\Varid{x}\mathrel{=}\mathbf{let}\;(\anonymous ,\Varid{c'})\mathrel{=}\Varid{ccoalg}\;\Varid{x}\;\mathbf{in}\;\Varid{c'}{}\<[E]%
\\[\blanklineskip]%
\>[B]{}\ceilingCoPointed{\cdot }\mathbin{::}(\Conid{Functor}\;\Varid{g})\Rightarrow (\Varid{a}\to \Varid{g}\;\Varid{a})\to \Varid{a}\to \copointed{\Varid{g}}\;\Varid{a}{}\<[E]%
\\
\>[B]{}\ceilingCoPointed{\Varid{g}}\mathrel{=}\Varid{id}\wedge\Varid{g}{}\<[E]%
\ColumnHook
\end{hscode}\resethooks
where \ensuremath{\wedge} is
\begin{hscode}\SaveRestoreHook
\column{B}{@{}>{\hspre}l<{\hspost}@{}}%
\column{E}{@{}>{\hspre}l<{\hspost}@{}}%
\>[B]{}(\wedge)\mathbin{::}(\Varid{x}\to \Varid{a})\to (\Varid{x}\to \Varid{b})\to (\Varid{x}\to (\Varid{a},\Varid{b})){}\<[E]%
\\
\>[B]{}(\Varid{f}\wedge\Varid{g})\;\Varid{x}\mathrel{=}(\Varid{f}\;\Varid{x},\Varid{g}\;\Varid{x}){}\<[E]%
\ColumnHook
\end{hscode}\resethooks

Now we are looking at repeating the development that we did in
\autoref{sec:simple-sos}. To do this, we will first need to turn the $\rho$-rule
into a distributive law. However, to be able to model more complex languages we
will need what have been called GSOS (for Guarded Structural Operational
Semantics) which subsume the copointed $\rho$-rules. Therefore, we will
introduce GSOS before giving the general method for constructing the semantics.

\section{Full Abstraction for CCS}
\label{sec:ccs}\label{sec:free-laws}
As a bigger case study we implement a fully abstract semantics for Milner's
\emph{Calculus of Communicating Systems} (CCS)~\citep{Milner80}. We first recap
CCS (\autoref{sec:ccs:recap}). We then explain our choice of the behavior
functor (\autoref{sec:choice-of-B}) and encode CCS' operational rules as a
$\rho$-rule (\autoref{sec:ccs:attempt}), resulting in a fully abstract semantics
for CCS, and finally, in \autoref{sec:ccs:examples} we demonstrate how the semantics can be used for stream programming through some programming examples.

\subsection{Syntax and Operational Rules of CCS}\label{sec:ccs:recap}
The syntax of value-passing CCS is as follows (relabelling and conditionals of
CCS~\citep{Milner80} are omitted for simplicity, but they can be
straightforwardly included):
\begin{equation}\label{eq:ccs-language}
  P := \overline{c}(v). P \mid c(x). P_x \mid P_1 + P_2 \mid \nil \mid P_1 \parallel P_2 \mid \rep(P) \mid P \restr c
\end{equation}
A CCS program, conventionally called a \emph{process}, $\overline{c}(v). P$
\emph{sends} a value $v \in \Val$ to the channel $c \in \Chan$ from some fixed
set $\Chan$ of channels, and continues as $P$.  Similarly, for a family of
processes $\{P_x\}_{x \in \Val}$, the process $c(x). P_x$ receives a
value---bound to the variable $x$---from the channel $c \in \Chan$, and
continues as $P_x$.  The \emph{sum} $P_1 + P_2$ of $P_1$ and $P_2$ can act as
both $P_1$ and $P_2$,  and the inactive process $\nil$ does nothing.  The
\emph{parallel composition} $P_1 \parallel P_2$ act as $P_1$ and $P_2$
concurrently, possibly communicating with each other.  The process $\rep(P)$ is
the \emph{replication} of arbitrarily many copies of $P$ in parallel.  Finally,
$P \restr c$ \emph{restricts} the communication on channel $c$ between $P$ and
the ambient environment.

The (small step) operational semantics of CCS is inductively defined in
\autoref{fig:ccs} as as a ternary relation
$(\to) \subseteq P \times \Act \times P$, where $\Act$ is the following set of
\emph{actions}:
\begin{equation}
\Act = \{\overline{c}(v) \mid c \in \Chan, v \in \ensuremath{\mathbb{V}} \} \cup \{c(v) \mid c \in \Chan, v \in \Val\} \cup \{\tau\}
\end{equation}
The intuition is that $P \reduceto{\overline{c}(v)} Q$ means that $P$ sends
a value $v$ to channel $c$ and becomes $Q$;
similarly $P \reduceto{c(v)} Q$ stands for $P$ receiving $v$ and becoming
$Q$; and finally there is a special action $\tau$, called the \emph{silent}
action, for which $P \reduceto{\tau} Q$ means that some communication
succeeds inside $P$, which is not visible to the ambient environment.

\begin{figure}
  \begin{mathpar}
  \inferrule { }{\overline{c}(v). P \reduceto{\overline{c}(v)} P}
  \and 
  \inferrule {v \in \Val }{c(x). P_x \reduceto{c(v)} P_{v}}
  \and 
  \inferrule {i \in \{0, 1\} \\ P_{i} \reduceto{a} P}{P_{0} + P_{1} \reduceto{a} P}
  \and
  \inferrule {P \reduceto{a} P'}{P \parallel Q \reduceto{a} P' \parallel Q}
  \and
  \inferrule {Q \reduceto{a} Q'}{P \parallel Q \reduceto{a} P\parallel Q'}
  \and
  \inferrule {P \reduceto{c(v)} P' \\ Q \reduceto{\overline{c}(v)} Q'}{P \parallel Q \reduceto{\tau} P' \parallel Q'}
  \and
  \inferrule {P \reduceto{\overline{c}(v)} P' \\ Q \reduceto{{c}(v)} Q'}{P \parallel Q \reduceto{\tau} P' \parallel Q'}
  \and
  \inferrule { P \reduceto{a} P' }{ \rep(P) \reduceto{a} P' \parallel \rep(P)}
  \and
  \inferrule {P \reduceto{a} P' \\ a \not\in \{c(v), \overline{c}(v)\}}{P \restr c \reduceto{a} P' \restr c}
\end{mathpar}\caption{Operational semantics for CCS}\label{fig:ccs}
\end{figure}

\subsection{The choice of the behavior functor}
\label{sec:choice-of-B}
The non-deterministic nature of the operational semantics in \autoref{fig:ccs}
is due to the non-deterministic ($P + Q$) and parallel ($P \parallel Q$)
operators. Thus, we need to use the finite powerset functor $\finpow{-}$ to
model transitions to a finite set of programs as we did in the previous section.

The reason we used the \emph{finite} powerset functor (or the finite lists over
a type) is rather theoretical: there is no fixed-point for the powerset functor
$\pow{-}$ which we need to construct the codomain for the interpretation
function\footnote[1]{This result is due to Cantor who proved that for every set
$X$, $X \not\iso \pow{X}$.}.

Moreover, we also need to capture the action coming from the transition
rules of the form  $P \reduceto{a} Q$.  Thus, the obvious choice for $B$ would be
$\finpow{\Act \times {-}}$.

However, consider the process $c(x). P_x$ receiving a value. Its possible
transitions $\{(c(v), P_v) \mid v \in \ensuremath{\mathbb{V}} \}$ are an \emph{infinite} subset of
$\Act \times P$. However, $\finpow{-}$ is the \emph{finite} subset functor and
it is thus not suitable for our purposes.

To overcome this problem we consider the following behavioral functor instead:
\[
  \finpow{ (\Chan \times \Val \times -) + (\Chan \times (-)^\Val) + 1 }
\]
Intuitively, an observable behavior has three possible shapes given by the
coproduct. A ``send'' rule produces a channel-value pair along with the target
process ($\Chan \times \Val \times X$). A ``receive'' rule produces a channel
together with a \emph{function} mapping each value $v$ to its process $P_{v}$
($\Chan \times X^\Val)$). Finally, a ``synchronisation'' between two processes
is presented by the type $1$.

\subsection{The choice of the rule format}\label{sec:ccs:attempt}
Now we need to encode the operational rules in \autoref{fig:ccs} in a
$\rho$-rule as we did in the previous sections. However, neither the simple rule
format  (\ref{rule:simple}) nor the copointed one (\ref{rule:copointed}) here
are suitable for implementing the rules of CCS. In particular, a limitation of
both rule formats is that the target in the conclusion must have \emph{exactly
one} constructor $\sigma'$ applied to the meta-variables, rather than multiple
nested constructors.

The operational rules of arithmetic language in \autoref{sec:simple-sos} and
\autoref{sec:copointed} happen to meet this requirement, but many rules of CCS
in \autoref{fig:ccs} do not.  For example, the rule for
$\overline{c}(v). P \reduceto{\overline{c}(v)} P$ has no constructors
applied to $P$, but the rule for
$ \rep(P) \reduceto{a} P' \parallel \rep(P)$ has two constructors applied to
$P$ and $P'$, namely $\parallel$ and $\rep$.

To overcome this limitation, we need to generalise the type of $\rho$-rules once again by allowing
arbitrarily many constructors in the target of transitions. We do this by
replacing \ensuremath{\Sigma} in the target of \ensuremath{\rho} with the \emph{free monad} over the
signature \ensuremath{\Sigma}. Ultimately, the type signature of $\rho$ should be as follows:
\begin{hscode}\SaveRestoreHook
\column{B}{@{}>{\hspre}l<{\hspost}@{}}%
\column{E}{@{}>{\hspre}l<{\hspost}@{}}%
\>[B]{}\rho\mathbin{::}\forall \Varid{x}\hsforall \hsdot{\cdot}{.\ }\Sigma\;(\copointed{\Varid{b}}\;\Varid{x})\to \Varid{b}\;(\free{\Sigma}\;\Varid{x}){}\<[E]%
\ColumnHook
\end{hscode}\resethooks
where \ensuremath{\free{\Sigma}} is the free monad. We now briefly recall its operations.

\subsubsection{Free Monads}
Given a functor $\Sigma$, the free monad $\Sigma^*$ over $\Sigma$ is defined by
the following datatype:
\begin{hscode}\SaveRestoreHook
\column{B}{@{}>{\hspre}l<{\hspost}@{}}%
\column{E}{@{}>{\hspre}l<{\hspost}@{}}%
\>[B]{}\mathbf{data}\;\free{\Sigma}\;\Varid{a}\mathrel{=}\Conid{Var}\;\Varid{a}\mid \Conid{Op}\;(\Sigma\;(\free{\Sigma}\;\Varid{a})){}\<[E]%
\ColumnHook
\end{hscode}\resethooks
which comes with the following recursion scheme:
\begin{hscode}\SaveRestoreHook
\column{B}{@{}>{\hspre}l<{\hspost}@{}}%
\column{23}{@{}>{\hspre}l<{\hspost}@{}}%
\column{E}{@{}>{\hspre}l<{\hspost}@{}}%
\>[B]{}\Varid{eval}\mathbin{::}\Conid{Functor}\;\Sigma\Rightarrow (\Sigma\;\Varid{b}\to \Varid{b})\to (\Varid{a}\to \Varid{b})\to \free{\Sigma}\;\Varid{a}\to \Varid{b}{}\<[E]%
\\
\>[B]{}\Varid{eval}\;\Varid{alg}\;\Varid{gen}\;(\Conid{Var}\;\Varid{v}){}\<[23]%
\>[23]{}\mathrel{=}\Varid{gen}\;\Varid{v}{}\<[E]%
\\
\>[B]{}\Varid{eval}\;\Varid{alg}\;\Varid{gen}\;(\Conid{Op}\;\Varid{c}){}\<[23]%
\>[23]{}\mathrel{=}\Varid{alg}\;(\Varid{fmap}\;(\Varid{eval}\;\Varid{alg}\;\Varid{gen})\;\Varid{c}){}\<[E]%
\ColumnHook
\end{hscode}\resethooks
Using \ensuremath{\Varid{eval}}, the monad instance of \ensuremath{\free{\Sigma}} is defined as
\begin{equation*}
\begin{codepage}[0.3\linewidth]
\setlength{\mathindent}{0cm}
\begin{hscode}\SaveRestoreHook
\column{B}{@{}>{\hspre}l<{\hspost}@{}}%
\column{E}{@{}>{\hspre}l<{\hspost}@{}}%
\>[B]{}\Varid{return}\mathbin{::}\Varid{a}\to \free{\Sigma}\;\Varid{a}{}\<[E]%
\\
\>[B]{}\Varid{return}\mathrel{=}\Conid{Var}{}\<[E]%
\ColumnHook
\end{hscode}\resethooks
\end{codepage}  
\begin{codepage}
\begin{hscode}\SaveRestoreHook
\column{B}{@{}>{\hspre}l<{\hspost}@{}}%
\column{E}{@{}>{\hspre}l<{\hspost}@{}}%
\>[B]{}(\bind )\mathbin{::}\Conid{Functor}\;\Sigma\Rightarrow \free{\Sigma}\;\Varid{a}\to (\Varid{a}\to \free{\Sigma}\;\Varid{b})\to \free{\Sigma}\;\Varid{b}{}\<[E]%
\\
\>[B]{}\Varid{x}\bind \Varid{f}\mathrel{=}\Varid{eval}\;\Conid{Op}\;\Varid{f}\;\Varid{x}{}\<[E]%
\ColumnHook
\end{hscode}\resethooks
\end{codepage}  
\end{equation*}

The recursion scheme above is justified by the fact that the free monad over a
signature functor \ensuremath{\Sigma} is equivalent to the least-fixed point of
$\Gamma X = A + \Sigma X$ satisfying $\mu \Gamma = A + \Sigma (\mu \Gamma)$.%

For every $\Sigma$-algebra we can construct a $\Sigma^{*}$-algebra by evaluating
the syntax using the identity function and the \ensuremath{\Sigma}-algebra and viceversa
\begin{equation*}
\begin{codepage}[0.51\linewidth]
\setlength{\mathindent}{0cm}
\begin{hscode}\SaveRestoreHook
\column{B}{@{}>{\hspre}l<{\hspost}@{}}%
\column{E}{@{}>{\hspre}l<{\hspost}@{}}%
\>[B]{}\floorFree{\cdot }\mathbin{::}\Conid{Functor}\;\Sigma\Rightarrow (\free{\Sigma}\;\Varid{a}\to \Varid{a})\to \Sigma\;\Varid{a}\to \Varid{a}{}\<[E]%
\\
\>[B]{}\floorFree{\Varid{f}}\mathrel{=}\Varid{f}\hsdot{\cdot}{.\ }\Conid{Op}\hsdot{\cdot}{.\ }\Varid{fmap}\;\Conid{Var}{}\<[E]%
\ColumnHook
\end{hscode}\resethooks
\end{codepage}
\begin{codepage}
\begin{hscode}\SaveRestoreHook
\column{B}{@{}>{\hspre}l<{\hspost}@{}}%
\column{E}{@{}>{\hspre}l<{\hspost}@{}}%
\>[B]{}\ceilingFree{\cdot }\mathbin{::}\Conid{Functor}\;\Sigma\Rightarrow (\Sigma\;\Varid{a}\to \Varid{a})\to \free{\Sigma}\;\Varid{a}\to \Varid{a}{}\<[E]%
\\
\>[B]{}\ceilingFree{\Varid{g}}\mathrel{=}\Varid{eval}\;\Varid{g}\;\Varid{id}{}\<[E]%
\ColumnHook
\end{hscode}\resethooks
\end{codepage}
\end{equation*}
The maps \ensuremath{\ceilingFree{\cdot }} and \ensuremath{\floorFree{\cdot }} witness the fact that $\Sigma$-algebras
and the $\Sigma^{*}$-algebras for the monad $\Sigma^{*}$ are in one-to-one correspondence\footnote[1]{Here we
mean that algebras \emph{for a monad} $\Sigma^{*}$ are isomorphic to algebras
for $\Sigma$}.

\subsubsection{The $\rho$-rule for CCS}
\label{sec:ccs:rho-rule}
As a first step we need to implement the syntax of the language using a
signature functor. This will be parameterised by a type \ensuremath{\mathbb{C}} of channel
labels.  As usual, \ensuremath{\mu(\Conid{CCS}\;\Varid{l})} will be the type of inductively generated CCS
programs (\ref{eq:ccs-language}).
\begin{hscode}\SaveRestoreHook
\column{B}{@{}>{\hspre}l<{\hspost}@{}}%
\column{13}{@{}>{\hspre}c<{\hspost}@{}}%
\column{13E}{@{}l@{}}%
\column{16}{@{}>{\hspre}l<{\hspost}@{}}%
\column{39}{@{}>{\hspre}l<{\hspost}@{}}%
\column{56}{@{}>{\hspre}l<{\hspost}@{}}%
\column{E}{@{}>{\hspre}l<{\hspost}@{}}%
\>[B]{}\mathbf{type}\;\mathbb{V}\mathrel{=}\Nat{}\<[E]%
\\
\>[B]{}\mathbf{type}\;\mathbb{C}\mathrel{=}\Conid{String}{}\<[E]%
\\[\blanklineskip]%
\>[B]{}\mathbf{data}\;\Conid{CCS}\;\Varid{x}{}\<[13]%
\>[13]{}\mathrel{=}{}\<[13E]%
\>[16]{}\Conid{Send}\;\mathbb{C}\;\mathbb{V}\;\Varid{x}\mid \Conid{Recv}\;\mathbb{C}\;(\mathbb{V}\to \Varid{x}){}\<[56]%
\>[56]{}\mid \Conid{Sum}\;\Varid{x}\;\Varid{x}{}\<[E]%
\\
\>[13]{}\mid {}\<[13E]%
\>[16]{}\Conid{Nil}\mid \Conid{Par}\;\Varid{x}\;\Varid{x}\mid \Conid{Rep}\;\Varid{x}{}\<[39]%
\>[39]{}\mid \Conid{Restrict}\;\mathbb{C}\;\Varid{x}{}\<[E]%
\ColumnHook
\end{hscode}\resethooks
Secondly, we need to define the set of actions
\begin{hscode}\SaveRestoreHook
\column{B}{@{}>{\hspre}l<{\hspost}@{}}%
\column{E}{@{}>{\hspre}l<{\hspost}@{}}%
\>[B]{}\mathbf{data}\;\Conid{Act}\;\Varid{x}\mathrel{=}\Conid{ActS}\;\mathbb{C}\;\mathbb{V}\;\Varid{x}\mid \Conid{ActR}\;\mathbb{C}\;(\mathbb{V}\to \Varid{x})\mid \Conid{Silent}\;\Varid{x}{}\<[E]%
\\[\blanklineskip]%
\>[B]{}\mathbf{newtype}\;\Conid{Acts}\;\Varid{x}\mathrel{=}\Conid{Acts}\;\{\mskip1.5mu \Varid{unActs}\mathbin{::}[\mskip1.5mu \Conid{Act}\;\Varid{x}\mskip1.5mu]\mskip1.5mu\}{}\<[E]%
\ColumnHook
\end{hscode}\resethooks
The following two definitions are simply the liftings of concatentation (\ensuremath{\mathbin{+\!\!+}})
and \ensuremath{\Varid{filter}} respectively to the type \ensuremath{\Conid{Acts}}
\begin{hscode}\SaveRestoreHook
\column{B}{@{}>{\hspre}l<{\hspost}@{}}%
\column{E}{@{}>{\hspre}l<{\hspost}@{}}%
\>[B]{}\Varid{bapp}\mathbin{::}\Conid{Acts}\;\Varid{x}\to \Conid{Acts}\;\Varid{x}\to \Conid{Acts}\;\Varid{x}{}\<[E]%
\\
\>[B]{}\Varid{bapp}\;\Varid{bs}\;\Varid{bs'}\mathrel{=}\Conid{Acts}\;(\Varid{unActs}\;\Varid{bs}\mathbin{+\!\!+}\Varid{unActs}\;\Varid{bs'}){}\<[E]%
\\[\blanklineskip]%
\>[B]{}\Varid{bfilter}\mathbin{::}(\Conid{Act}\;\Varid{x}\to 2)\to \Conid{Acts}\;\Varid{x}\to \Conid{Acts}\;\Varid{x}{}\<[E]%
\\
\>[B]{}\Varid{bfilter}\;\Varid{p}\mathrel{=}\Conid{Acts}\hsdot{\cdot}{.\ }\Varid{filter}\;\Varid{p}\hsdot{\cdot}{.\ }\Varid{unActs}{}\<[E]%
\ColumnHook
\end{hscode}\resethooks

Now the operational rules of CCS in \autoref{fig:ccs} can be encoded as a $\rho$-rule of type
\begin{hscode}\SaveRestoreHook
\column{B}{@{}>{\hspre}l<{\hspost}@{}}%
\column{12}{@{}>{\hspre}l<{\hspost}@{}}%
\column{E}{@{}>{\hspre}l<{\hspost}@{}}%
\>[B]{}\rho_{\Varid{CCS}}\mathbin{::}{}\<[12]%
\>[12]{}\Conid{CCS}\;(\Varid{x},\Conid{Acts}\;\Varid{x})\to \Conid{Acts}\;(\free{\Conid{CCS}}\;\Varid{x}){}\<[E]%
\ColumnHook
\end{hscode}\resethooks
The rules for sending and receiving are as follows
\begin{hscode}\SaveRestoreHook
\column{B}{@{}>{\hspre}l<{\hspost}@{}}%
\column{30}{@{}>{\hspre}l<{\hspost}@{}}%
\column{E}{@{}>{\hspre}l<{\hspost}@{}}%
\>[B]{}\rho_{\Varid{CCS}}\;(\Conid{Send}\;\Varid{c}\;\Varid{v}\;(\Varid{x},\anonymous )){}\<[30]%
\>[30]{}\mathrel{=}\Conid{Acts}\;[\mskip1.5mu \Conid{ActS}\;\Varid{c}\;\Varid{v}\;(\Varid{return}\;\Varid{x})\mskip1.5mu]{}\<[E]%
\\
\>[B]{}\rho_{\Varid{CCS}}\;(\Conid{Recv}\;\Varid{c}\;\Varid{k}){}\<[30]%
\>[30]{}\mathrel{=}\Conid{Acts}\;[\mskip1.5mu \Conid{ActR}\;\Varid{c}\;(\Varid{fmap}\;(\lambda (\Varid{x},\anonymous )\to \Varid{return}\;\Varid{x})\;\Varid{k})\mskip1.5mu]{}\<[E]%
\ColumnHook
\end{hscode}\resethooks
where the sending process has one single action, i.e. the send action along with the channel, the value and the copy of the remaining computation and the receiving process has one single action
as well, i.e. the receive action with the channel and the continuation that
always return the process after receiving the value $v$.

The case of the $0$ (nil) process is just the empty set of actions
\begin{hscode}\SaveRestoreHook
\column{B}{@{}>{\hspre}l<{\hspost}@{}}%
\column{30}{@{}>{\hspre}l<{\hspost}@{}}%
\column{E}{@{}>{\hspre}l<{\hspost}@{}}%
\>[B]{}\rho_{\Varid{CCS}}\;\Conid{Nil}{}\<[30]%
\>[30]{}\mathrel{=}\Conid{Acts}\;[\mskip1.5mu \mskip1.5mu]{}\<[E]%
\ColumnHook
\end{hscode}\resethooks

The choice process ($P_{0} + P_{1}$) is the concatenation of the list of actions from the first
process with the second process, while the restriction ($P \backslash c$) requires some additional
functions. The first is \ensuremath{\Varid{check}} implementing the condition $a \not\in \{c(v), \overline{c}(v)\}$ (\autoref{fig:ccs})
and the other is \ensuremath{\Varid{bfilter}} filtering out all the actions that satisfies the
condition implemented by \ensuremath{\Varid{check}}
\begin{hscode}\SaveRestoreHook
\column{B}{@{}>{\hspre}l<{\hspost}@{}}%
\column{3}{@{}>{\hspre}l<{\hspost}@{}}%
\column{24}{@{}>{\hspre}l<{\hspost}@{}}%
\column{30}{@{}>{\hspre}l<{\hspost}@{}}%
\column{E}{@{}>{\hspre}l<{\hspost}@{}}%
\>[B]{}\rho_{\Varid{CCS}}\;(\Conid{Sum}\;(\anonymous ,\Varid{b})\;(\anonymous ,\Varid{b'})){}\<[30]%
\>[30]{}\mathrel{=}\Varid{fmap}\;\Varid{return}\;\Varid{b}\mathbin{+\!\!+_{\Varid{B}}}\Varid{fmap}\;\Varid{return}\;\Varid{b'}{}\<[E]%
\\
\>[B]{}\rho_{\Varid{CCS}}\;(\Conid{Restrict}\;\Varid{c}\;(\anonymous ,\Varid{b})){}\<[30]%
\>[30]{}\mathrel{=}\Varid{bfilter}\;\Varid{check}\;(\Varid{fmap}\;(\lambda \Varid{y}\to \Conid{Op}\;(\Conid{Restrict}\;\Varid{c}\;(\Varid{return}\;\Varid{y})))\;\Varid{b})\;\mathbf{where}{}\<[E]%
\\
\>[B]{}\hsindent{3}{}\<[3]%
\>[3]{}\Varid{check}\;(\Conid{ActS}\;\Varid{c'}\;\anonymous \;\anonymous ){}\<[24]%
\>[24]{}\mathrel{=}\Varid{c}\not\equiv \Varid{c'}{}\<[E]%
\\
\>[B]{}\hsindent{3}{}\<[3]%
\>[3]{}\Varid{check}\;(\Conid{ActR}\;\Varid{c'}\;\anonymous ){}\<[24]%
\>[24]{}\mathrel{=}\Varid{c}\not\equiv \Varid{c'}{}\<[E]%
\\
\>[B]{}\hsindent{3}{}\<[3]%
\>[3]{}\Varid{check}\;(\Conid{Silent}\;\anonymous ){}\<[24]%
\>[24]{}\mathrel{=}\Conid{True}{}\<[E]%
\ColumnHook
\end{hscode}\resethooks

The parallel case ($\parallel$) implements four rules in one so it is perhaps not surprising that its
definition is a bit more involved
\begin{hscode}\SaveRestoreHook
\column{B}{@{}>{\hspre}l<{\hspost}@{}}%
\column{3}{@{}>{\hspre}l<{\hspost}@{}}%
\column{32}{@{}>{\hspre}l<{\hspost}@{}}%
\column{37}{@{}>{\hspre}l<{\hspost}@{}}%
\column{43}{@{}>{\hspre}l<{\hspost}@{}}%
\column{E}{@{}>{\hspre}l<{\hspost}@{}}%
\>[B]{}\rho_{\Varid{CCS}}\;(\Conid{Par}\;\Varid{xb}\;\Varid{xb'}){}\<[32]%
\>[32]{}\mathrel{=}\Varid{lmerge}\;\Varid{xb}\;\Varid{xb'}\mathbin{+\!\!+_{\Varid{B}}}\Varid{lmerge}\;\Varid{xb'}\;\Varid{xb}{}\<[E]%
\\
\>[B]{}\hsindent{3}{}\<[3]%
\>[3]{}\mathbf{where}\;\Varid{lmerge}\;(\Varid{x},\Varid{b})\;(\Varid{x'},\Varid{b'})\mathrel{=}\Varid{fmap}\;(\lambda \Varid{y}\to \Conid{Op}\;(\Conid{Par}\;(\Varid{return}\;\Varid{y})\;(\Varid{return}\;\Varid{x'})))\;\Varid{b}\mathbin{+\!\!+_{\Varid{B}}}{}\<[E]%
\\
\>[3]{}\hsindent{34}{}\<[37]%
\>[37]{}\Conid{Acts}\;[\mskip1.5mu \Conid{Silent}\;(\Conid{Op}\;(\Conid{Par}\;(\Varid{return}\;\Varid{y})\;(\Varid{return}\;(\Varid{k}\;\Varid{m})))){}\<[E]%
\\
\>[37]{}\hsindent{6}{}\<[43]%
\>[43]{}\mid (\Conid{ActS}\;\Varid{c}\;\Varid{m}\;\Varid{y})\leftarrow \Varid{unActs}\;\Varid{b},(\Conid{ActR}\;\Varid{c'}\;\Varid{k})\leftarrow \Varid{unActs}\;\Varid{b'},\Varid{c}\equiv \Varid{c'}\mskip1.5mu]{}\<[E]%
\ColumnHook
\end{hscode}\resethooks
here \ensuremath{\Varid{lmerge}} takes the initial abstract processes \ensuremath{\Varid{x}} and \ensuremath{\Varid{x'}} and their
respective list of actions \ensuremath{\Varid{b}} and \ensuremath{\Varid{b'}} and it produces the union of the list of
actions coming from a one-step reduction of \ensuremath{\Varid{x}} and the list of silent actions
coming from the coupling of a send from \ensuremath{\Varid{x}} and a receive from \ensuremath{\Varid{x'}}. At the top
level, the rule outputs the union of the results of merging the outputs of  \ensuremath{\Varid{x}}
and \ensuremath{\Varid{x'}} and their mirror case.

The rule for $\rep$ is simpler to implement but crucially requires the structure
from the free monad as it returns two levels of syntax, the first is the \ensuremath{\Conid{Par}}
constructor and the second constructor is \ensuremath{\Conid{Rep}}
\begin{hscode}\SaveRestoreHook
\column{B}{@{}>{\hspre}l<{\hspost}@{}}%
\column{E}{@{}>{\hspre}l<{\hspost}@{}}%
\>[B]{}\rho_{\Varid{CCS}}\;(\Conid{Rep}\;(\Varid{x},\Varid{b}))\mathrel{=}\Varid{fmap}\;(\lambda \Varid{y}\to \Conid{Op}\;(\Conid{Par}\;(\Varid{return}\;\Varid{y})\;(\Conid{Op}\;(\Conid{Rep}\;(\Varid{return}\;\Varid{x})))))\;\Varid{b}{}\<[E]%
\ColumnHook
\end{hscode}\resethooks

At this point, we need to recover the distributive law for GSOS rules from which
we can construct the universal semantics.

\subsection{The general case: recovering distributive laws}
The $\rho$-rule for CCS can be generalised in the following way.  For a
signature functor \ensuremath{\Sigma} and a behavior functor \ensuremath{\Varid{b}} the $\rho$-rule has the
following type:
\begin{hscode}\SaveRestoreHook
\column{B}{@{}>{\hspre}l<{\hspost}@{}}%
\column{E}{@{}>{\hspre}l<{\hspost}@{}}%
\>[B]{}\rho\mathbin{::}\forall \Varid{x}\hsforall \hsdot{\cdot}{.\ }\Sigma\;(\copointed{\Varid{b}}\;\Varid{x})\to \free{\Sigma}\;(\Varid{b}\;\Varid{x}){}\<[E]%
\ColumnHook
\end{hscode}\resethooks

This parametric function gives raise to a rule format called GSOS
\begin{equation}\label{rule:copointed}
\inferrule{x_{1}\reduceto{l_{1}} x'_{1} \dots x_{n} \reduceto{l_{n}} x'_{n}}{\sigma(x_1, \cdots, x_n) \reduceto{a} t(y_1, \cdots, y_m)}
\end{equation}
for meta-variables $\{y_{1} \dots, y_{m}\} \subseteq \{x_{1}, \dots, x_{n}\} \cup \{x'_{1}, \dots, x'_{n}\}$ and constructors $\sigma, \in \Sigma$ and programs $t \in \Sigma^{*}$.

Now we need to recover a distributive law for the free monad \ensuremath{\free{\Sigma}} over
\ensuremath{\copointed{\Conid{B}}} from \ensuremath{\rho}. We can do this by using the recursion scheme for the free monad \ensuremath{\Varid{eval}}:
\begin{hscode}\SaveRestoreHook
\column{B}{@{}>{\hspre}l<{\hspost}@{}}%
\column{3}{@{}>{\hspre}l<{\hspost}@{}}%
\column{10}{@{}>{\hspre}l<{\hspost}@{}}%
\column{60}{@{}>{\hspre}l<{\hspost}@{}}%
\column{E}{@{}>{\hspre}l<{\hspost}@{}}%
\>[B]{}\Varid{rhoToLambda}\mathbin{::}\forall \Sigma\hsforall \;\Varid{b}\hsdot{\cdot}{.\ }(\Conid{Functor}\;\Sigma,\Conid{Functor}\;\Varid{b}){}\<[60]%
\>[60]{}\Rightarrow (\forall \Varid{x}\hsforall \hsdot{\cdot}{.\ }\Sigma\;(\copointed{\Varid{b}}\;\Varid{x})\to \Varid{b}\;(\free{\Sigma}\;\Varid{x})){}\<[E]%
\\
\>[60]{}\to (\forall \Varid{x}\hsforall \hsdot{\cdot}{.\ }\free{\Sigma}\;(\copointed{\Varid{b}}\;\Varid{x})\to \copointed{\Varid{b}}\;(\free{\Sigma}\;\Varid{x})){}\<[E]%
\\
\>[B]{}\Varid{rhoToLambda}\;\rho\mathrel{=}\Varid{eval}\;\Varid{alg}\;(\Varid{pmap}\;\Varid{return})\;\mathbf{where}{}\<[E]%
\\
\>[B]{}\hsindent{3}{}\<[3]%
\>[3]{}\Varid{alg}\mathbin{::}\Sigma\;(\copointed{\Varid{b}}\;(\free{\Sigma}\;\Varid{x}))\to \copointed{\Varid{b}}\;(\free{\Sigma}\;\Varid{x}){}\<[E]%
\\
\>[B]{}\hsindent{3}{}\<[3]%
\>[3]{}\Varid{alg}\mathrel{=}{}\<[10]%
\>[10]{}(\Conid{Op}\hsdot{\cdot}{.\ }(\Varid{fmap}\;\epsilon))\wedge((\Varid{fmap}\;(\bind \Varid{id}))\hsdot{\cdot}{.\ }\rho){}\<[E]%
\ColumnHook
\end{hscode}\resethooks
With the distributive law \ensuremath{\Varid{rhoToLambda}\;\rho}, the universal operational \ensuremath{\Varid{opsem}}
and denotational semantics \ensuremath{\Varid{desem}} induced by \ensuremath{\rho} can be defined similarly to
what we did in the previous sections. Only here we need to lift and downlift the
algebras using the operators \ensuremath{\ceilingFree{\cdot }}, \ensuremath{\floorFree{\cdot }}, \ensuremath{\ceilingCoPointed{\cdot }} and
\ensuremath{\floorCoPointed{\cdot }} to make the types match
\begin{hscode}\SaveRestoreHook
\column{B}{@{}>{\hspre}l<{\hspost}@{}}%
\column{3}{@{}>{\hspre}l<{\hspost}@{}}%
\column{114}{@{}>{\hspre}l<{\hspost}@{}}%
\column{E}{@{}>{\hspre}l<{\hspost}@{}}%
\>[B]{}\Varid{opsem}\mathbin{::}\forall \Sigma\hsforall \;\Varid{b}\hsdot{\cdot}{.\ }(\Conid{Functor}\;\Sigma,\Conid{Functor}\;\Varid{b})\Rightarrow (\forall \Varid{x}\hsforall \hsdot{\cdot}{.\ }\Sigma\;(\copointed{\Varid{b}}\;\Varid{x})\to \Varid{b}\;(\free{\Sigma}\;\Varid{x}))\to {}\<[114]%
\>[114]{}\mu\Sigma\to \Varid{b}\;(\mu\Sigma){}\<[E]%
\\
\>[B]{}\Varid{opsem}\;\rho\mathrel{=}\floorCoPointed{\catamor{\Varid{alg}}}\;\mathbf{where}{}\<[E]%
\\
\>[B]{}\hsindent{3}{}\<[3]%
\>[3]{}\Varid{alg}\mathbin{::}\Sigma\;(\copointed{\Varid{b}}\;(\mu\Sigma))\to \copointed{\Varid{b}}\;(\mu\Sigma){}\<[E]%
\\
\>[B]{}\hsindent{3}{}\<[3]%
\>[3]{}\Varid{alg}\mathrel{=}\floorFree{\Varid{pmap}\;(\ceilingFree{\Conid{In}})\hsdot{\cdot}{.\ }(\Varid{rhoToLambda}\;\rho)}{}\<[E]%
\ColumnHook
\end{hscode}\resethooks
and similarly for the denotational semantics:
\begin{hscode}\SaveRestoreHook
\column{B}{@{}>{\hspre}l<{\hspost}@{}}%
\column{3}{@{}>{\hspre}l<{\hspost}@{}}%
\column{E}{@{}>{\hspre}l<{\hspost}@{}}%
\>[B]{}\Varid{desem}\mathbin{::}\forall \Sigma\hsforall \;\Varid{b}\hsdot{\cdot}{.\ }(\Conid{Functor}\;\Sigma,\Conid{Functor}\;\Varid{b})\Rightarrow (\forall \Varid{x}\hsforall \hsdot{\cdot}{.\ }\Sigma\;(\copointed{\Varid{b}}\;\Varid{x})\to \Varid{b}\;(\free{\Sigma}\;\Varid{x}))\to \Sigma\;(\nu\Varid{b})\to \nu\Varid{b}{}\<[E]%
\\
\>[B]{}\Varid{desem}\;\rho\mathrel{=}\floorFree{\ana{\Varid{coalg}}}\;\mathbf{where}{}\<[E]%
\\
\>[B]{}\hsindent{3}{}\<[3]%
\>[3]{}\Varid{coalg}\mathbin{::}\free{\Sigma}\;(\nu\Varid{b})\to \Varid{b}\;(\free{\Sigma}\;(\nu\Varid{b})){}\<[E]%
\\
\>[B]{}\hsindent{3}{}\<[3]%
\>[3]{}\Varid{coalg}\mathrel{=}\floorCoPointed{\Varid{rhoToLambda}\;\rho\hsdot{\cdot}{.\ }\Varid{fmap}\;(\ceilingCoPointed{\Varid{out}})}{}\<[E]%
\ColumnHook
\end{hscode}\resethooks

\subsubsection{The Full Abstraction Argument}
Since \ensuremath{\lambda} is distributive, the functions \ensuremath{\catamor{\Varid{desem}\;\rho}} and \ensuremath{\ana{\Varid{opsem}\;\rho}} always coincide~\citep{HinzeJ11}, and, since \ensuremath{\ana{\Varid{opsem}\;\rho}}
is an unfold it is always fully abstract. The following recursion scheme
interprets a language \ensuremath{\mu\Sigma} provided an abstract rule \ensuremath{\rho} (as usual, we
write $\den{t}\rho$ for \ensuremath{\Varid{sem}\;\rho\;\Varid{t}}):
\begin{hscode}\SaveRestoreHook
\column{B}{@{}>{\hspre}l<{\hspost}@{}}%
\column{19}{@{}>{\hspre}l<{\hspost}@{}}%
\column{E}{@{}>{\hspre}l<{\hspost}@{}}%
\>[B]{}\Varid{sem}\mathbin{::}\forall \Sigma\hsforall \;\Varid{b}\hsdot{\cdot}{.\ }(\Conid{Functor}\;\Sigma,\Conid{Functor}\;\Varid{b})\Rightarrow (\forall \Varid{x}\hsforall \hsdot{\cdot}{.\ }\Sigma\;(\copointed{\Varid{b}}\;\Varid{x})\to \Varid{b}\;(\free{\Sigma}\;\Varid{x}))\to \mu\Sigma\to \nu\Varid{b}{}\<[E]%
\\
\>[B]{}\Varid{sem}\;\rho\;\Varid{t}\mathrel{=}{}\<[19]%
\>[19]{}\catamor{\Varid{desem}\;\rho}\;\Varid{t}\mbox{\onelinecomment  or $\ensuremath{\ana{\Varid{opsem}\;\rho}\;\Varid{t}}$}{}\<[E]%
\ColumnHook
\end{hscode}\resethooks

Applying \ensuremath{\Varid{sem}} to \ensuremath{\rho_{\Varid{CCS}}}, we obtain the following function that gives semantics to CCS programs as their (global) behaviours:
\begin{hscode}\SaveRestoreHook
\column{B}{@{}>{\hspre}l<{\hspost}@{}}%
\column{E}{@{}>{\hspre}l<{\hspost}@{}}%
\>[B]{}\Varid{sem_{CCS}}\mathbin{::}\mu\Conid{CCS}\to \nu\Conid{Acts}{}\<[E]%
\\
\>[B]{}\Varid{sem_{CCS}}\mathrel{=}\Varid{sem}\;\rho_{\Varid{CCS}}{}\<[E]%
\ColumnHook
\end{hscode}\resethooks

In the rest of this section, we continue our example of CCS with this recursion
scheme, and show how it be used for functional programming.

\subsection{Programming Examples}
\label{sec:ccs:examples}
In the rest of this section, we demonstrate how the semantics induced by
$\rho$-rules is useful through programming examples in CCS. For convenience, we
define the following helper functions wrapping constructors with \ensuremath{\Conid{In}}:
\begin{hscode}\SaveRestoreHook
\column{B}{@{}>{\hspre}l<{\hspost}@{}}%
\column{13}{@{}>{\hspre}l<{\hspost}@{}}%
\column{38}{@{}>{\hspre}l<{\hspost}@{}}%
\column{55}{@{}>{\hspre}l<{\hspost}@{}}%
\column{66}{@{}>{\hspre}l<{\hspost}@{}}%
\column{E}{@{}>{\hspre}l<{\hspost}@{}}%
\>[B]{}\Varid{p}\mathbin{\vert\vert}\Varid{q}{}\<[13]%
\>[13]{}\mathrel{=}\Conid{In}\;(\Conid{Par}\;\Varid{p}\;\Varid{q});{}\<[38]%
\>[38]{}\hspace{2em}\;\hspace{2em}\;{}\<[55]%
\>[55]{}\Varid{nil}{}\<[66]%
\>[66]{}\mathrel{=}\Conid{In}\;\Conid{Nil}{}\<[E]%
\\
\>[B]{}\Varid{send}\;\Varid{c}\;\Varid{m}\;\Varid{p}{}\<[13]%
\>[13]{}\mathrel{=}\Conid{In}\;(\Conid{Send}\;\Varid{c}\;\Varid{m}\;\Varid{p});{}\<[55]%
\>[55]{}\Varid{sum}\;\Varid{p}\;\Varid{q}{}\<[66]%
\>[66]{}\mathrel{=}\Conid{In}\;(\Conid{Sum}\;\Varid{p}\;\Varid{q}){}\<[E]%
\\
\>[B]{}\Varid{rep}\;\Varid{p}{}\<[13]%
\>[13]{}\mathrel{=}\Conid{In}\;(\Conid{Rep}\;\Varid{p});{}\<[55]%
\>[55]{}\Varid{res}\;\Varid{c}\;\Varid{p}{}\<[66]%
\>[66]{}\mathrel{=}\Conid{In}\;(\Conid{Restrict}\;\Varid{c}\;\Varid{p}){}\<[E]%
\\
\>[B]{}\Varid{recv}\;\Varid{c}\;\Varid{k}{}\<[13]%
\>[13]{}\mathrel{=}\Conid{In}\;(\Conid{Recv}\;\Varid{c}\;\Varid{k});{}\<[55]%
\>[55]{}\Varid{res'}\;\Varid{ls}\;\Varid{p}{}\<[66]%
\>[66]{}\mathrel{=}\Varid{foldr}\;\Varid{res}\;\Varid{p}\;\Varid{ls}{}\<[E]%
\ColumnHook
\end{hscode}\resethooks

These are just the constructors of the language post-applied to \ensuremath{\Conid{In}} while
\ensuremath{\Varid{res'}} is the restriction of a list of channels \ensuremath{\Varid{ls}} to the process \ensuremath{\Varid{p}}.

\begin{example}
We can use CCS to do stream programming.
The following program sends all natural numbers to the channel \ensuremath{\text{\ttfamily \char34 output\char34}} one
by one:
\begin{hscode}\SaveRestoreHook
\column{B}{@{}>{\hspre}l<{\hspost}@{}}%
\column{3}{@{}>{\hspre}l<{\hspost}@{}}%
\column{E}{@{}>{\hspre}l<{\hspost}@{}}%
\>[B]{}\Varid{nats}\mathbin{::}\mu\Conid{CCS}{}\<[E]%
\\
\>[B]{}\Varid{nats}\mathrel{=}\Varid{res}\;\text{\ttfamily \char34 i\char34}\;(\Varid{send}\;\text{\ttfamily \char34 i\char34}\;\mathrm{0}\;\Varid{nil}\mathbin{\vert\vert}\Varid{rep}\;\Varid{iter})\;\mathbf{where}{}\<[E]%
\\
\>[B]{}\hsindent{3}{}\<[3]%
\>[3]{}\Varid{iter}\mathrel{=}\Varid{recv}\;\text{\ttfamily \char34 i\char34}\;(\lambda \Varid{i}\to \Varid{send}\;\text{\ttfamily \char34 output\char34}\;\Varid{i}\;(\Varid{send}\;\text{\ttfamily \char34 i\char34}\;(\Varid{i}\mathbin{+}\mathrm{1})\;\Varid{nil})){}\<[E]%
\ColumnHook
\end{hscode}\resethooks
A process \ensuremath{\Varid{iter}} receives the current value from channel
\ensuremath{\text{\ttfamily \char34 i\char34}} and sending it to the \ensuremath{\text{\ttfamily \char34 output\char34}} channel.
The process \ensuremath{\Varid{iter}} is replicated as infinitely many copies by \ensuremath{\Varid{rep}}, so \ensuremath{\Varid{iter}}
can send \ensuremath{\Varid{i}\mathbin{+}\mathrm{1}} to another copy of itself to continue the iteration.
Finally, there is a \ensuremath{\Varid{send}\;\text{\ttfamily \char34 i\char34}\;\mathrm{0}\;\Varid{nil}} to start the iteration, and \ensuremath{\Varid{res}\;\text{\ttfamily \char34 i\char34}}
restricts the communication on \ensuremath{\text{\ttfamily \char34 i\char34}} inside the process, so other processes
cannot meddle with the iteration, and can only observe the channel \ensuremath{\text{\ttfamily \char34 output\char34}}.

Applying the semantics function to the process, we obtain \ensuremath{\Varid{sem_{CCS}}\;\Varid{nats}\mathbin{::}\nu\Conid{Acts}}, which is intuitively a (coinductive) tree whose edges are labelled
with an action \ensuremath{\Conid{Act}}, and different paths stand for different possibilities of
non-determinism.  We can collect all the outputs as a list in such a tree using
the following function:
\begin{hscode}\SaveRestoreHook
\column{B}{@{}>{\hspre}l<{\hspost}@{}}%
\column{3}{@{}>{\hspre}l<{\hspost}@{}}%
\column{21}{@{}>{\hspre}l<{\hspost}@{}}%
\column{E}{@{}>{\hspre}l<{\hspost}@{}}%
\>[B]{}\Varid{outputs}\mathbin{::}\nu\Conid{Acts}\to [\mskip1.5mu \mathbb{V}\mskip1.5mu]{}\<[E]%
\\
\>[B]{}\Varid{outputs}\;(\Conid{Out}^{\circ}\;(\Conid{Acts}\;[\mskip1.5mu \mskip1.5mu]))\mathrel{=}[\mskip1.5mu \mskip1.5mu]{}\<[E]%
\\
\>[B]{}\Varid{outputs}\;(\Conid{Out}^{\circ}\;(\Conid{Acts}\;\Varid{bs}))\mathrel{=}\Varid{concat}\;(\Varid{map}\;\Varid{f}\;\Varid{bs})\;\mathbf{where}{}\<[E]%
\\
\>[B]{}\hsindent{3}{}\<[3]%
\>[3]{}\Varid{f}\;(\Conid{ActS}\;\anonymous \;\Varid{m}\;\Varid{bs'}){}\<[21]%
\>[21]{}\mathrel{=}\Varid{m}\mathbin{:}\Varid{outputs}\;\Varid{bs'}{}\<[E]%
\\
\>[B]{}\hsindent{3}{}\<[3]%
\>[3]{}\Varid{f}\;(\Conid{ActR}\;\anonymous \;\anonymous ){}\<[21]%
\>[21]{}\mathrel{=}[\mskip1.5mu \mskip1.5mu]{}\<[E]%
\\
\>[B]{}\hsindent{3}{}\<[3]%
\>[3]{}\Varid{f}\;(\Conid{Silent}\;\Varid{bs'}){}\<[21]%
\>[21]{}\mathrel{=}\Varid{outputs}\;\Varid{bs'}{}\<[E]%
\ColumnHook
\end{hscode}\resethooks
And indeed, \ensuremath{\Varid{outputs}\;(\Varid{sem_{CCS}}\;\Varid{nats})\mathrel{=}[\mskip1.5mu \mathrm{0},\mathrm{1},\mathrm{2},\mathrm{3},\mathbin{...}\mskip1.5mu]}.
\end{example}

\begin{example}
The semantics \ensuremath{\Varid{sem_{CCS}}} is useful for exploring the non-deterministic behaviour
of concurrent systems, since all non-deterministic possibilities are recorded
in the semantics \ensuremath{\nu\Conid{Acts}}.
As an example, the following program implements a counter using the subprograms
defined at the beginning of this section:
\begin{hscode}\SaveRestoreHook
\column{B}{@{}>{\hspre}l<{\hspost}@{}}%
\column{3}{@{}>{\hspre}l<{\hspost}@{}}%
\column{17}{@{}>{\hspre}l<{\hspost}@{}}%
\column{E}{@{}>{\hspre}l<{\hspost}@{}}%
\>[B]{}\Varid{counter}\mathrel{=}\Varid{res}\;\text{\ttfamily \char34 init\char34}\;(\Varid{send}\;\text{\ttfamily \char34 init\char34}\;\mathrm{0}\;\Varid{nil}\mathbin{\vert\vert}\Varid{iter})\;\mathbf{where}{}\<[E]%
\\
\>[B]{}\hsindent{3}{}\<[3]%
\>[3]{}\Varid{iter}\mathrel{=}\Varid{rep}\;(\Varid{recv}\;\text{\ttfamily \char34 init\char34}\;(\lambda \Varid{v}\to \Varid{sum}{}\<[E]%
\\
\>[3]{}\hsindent{14}{}\<[17]%
\>[17]{}(\Varid{recv}\;\text{\ttfamily \char34 rd\char34}\;(\mathbin{\char92 \char95 }\to \Varid{send}\;\text{\ttfamily \char34 count\char34}\;\Varid{v}\;(\Varid{send}\;\text{\ttfamily \char34 init\char34}\;\Varid{v}\;\Varid{nil}))){}\<[E]%
\\
\>[3]{}\hsindent{14}{}\<[17]%
\>[17]{}(\Varid{recv}\;\text{\ttfamily \char34 wt\char34}\;(\lambda \Varid{i}\to \Varid{send}\;\text{\ttfamily \char34 init\char34}\;\Varid{i}\;\Varid{nil})))){}\<[E]%
\ColumnHook
\end{hscode}\resethooks
Intuitively, \ensuremath{\Varid{counter}} is a mobile program storing a value called \ensuremath{\Varid{init}} and
responding to read or write requests from other processes which would like to
read or overwrite the contents of the store.

The cell can be read and modified by accessing the channels \ensuremath{\text{\ttfamily \char34 rd\char34}}, \ensuremath{\text{\ttfamily \char34 wt\char34}}, and
\ensuremath{\text{\ttfamily \char34 count\char34}} as follows:
{
\setlength{\mathindent}{0cm}
\begin{hscode}\SaveRestoreHook
\column{B}{@{}>{\hspre}l<{\hspost}@{}}%
\column{46}{@{}>{\hspre}l<{\hspost}@{}}%
\column{55}{@{}>{\hspre}l<{\hspost}@{}}%
\column{75}{@{}>{\hspre}l<{\hspost}@{}}%
\column{E}{@{}>{\hspre}l<{\hspost}@{}}%
\>[B]{}\Varid{read}\mathbin{::}(\Nat\to \mu\Conid{CCS})\to \mu\Conid{CCS}\;{}\<[46]%
\>[46]{}\hspace{2em}\;{}\<[55]%
\>[55]{}\Varid{write}\mathbin{::}\Nat\to \mu\Conid{CCS}\to \mu\Conid{CCS}{}\<[E]%
\\
\>[B]{}\Varid{read}\;\Varid{k}\mathrel{=}\Varid{send}\;\text{\ttfamily \char34 rd\char34}\;\mathrm{0}\;(\Varid{recv}\;\text{\ttfamily \char34 count\char34}\;\Varid{k})\;{}\<[75]%
\>[75]{}\Varid{write}\;\Varid{v}\;\Varid{p}\mathrel{=}\Varid{send}\;\text{\ttfamily \char34 wt\char34}\;\Varid{v}\;\Varid{p}{}\<[E]%
\ColumnHook
\end{hscode}\resethooks
}
Suppose there are two processes incrementing the counter concurrently without
using a lock as shown in the following program:
\begin{hscode}\SaveRestoreHook
\column{B}{@{}>{\hspre}l<{\hspost}@{}}%
\column{3}{@{}>{\hspre}l<{\hspost}@{}}%
\column{E}{@{}>{\hspre}l<{\hspost}@{}}%
\>[B]{}\Varid{incr}\mathbin{::}\mu\Conid{CCS}\to \mu\Conid{CCS}{}\<[E]%
\\
\>[B]{}\Varid{incr}\;\Varid{p}\mathrel{=}\Varid{read}\;(\lambda \Varid{v}\to \Varid{write}\;(\Varid{v}\mathbin{+}\mathrm{1})\;\Varid{p}){}\<[E]%
\\[\blanklineskip]%
\>[B]{}\Varid{counterTest}\mathrel{=}\Varid{res'}\;[\mskip1.5mu \text{\ttfamily \char34 rd\char34},\text{\ttfamily \char34 wt\char34},\text{\ttfamily \char34 count\char34}\mskip1.5mu]\;(\Varid{counter}\mathbin{\vert\vert}\Varid{incr}\;(\Varid{incr}\;\Varid{nil}){}\<[E]%
\\
\>[B]{}\hsindent{3}{}\<[3]%
\>[3]{}\mathbin{\vert\vert}\Varid{incr}\;(\Varid{incr}\;(\Varid{read}\;(\lambda \Varid{v}\to \Varid{send}\;\text{\ttfamily \char34 output\char34}\;\Varid{v}\;\Varid{nil})))){}\<[E]%
\ColumnHook
\end{hscode}\resethooks
Then executing the semantics tells us all possible outcomes: 
\[\ensuremath{\Varid{nub}\;(\Varid{outputs}\;(\Varid{sem_{CCS}}\;\Varid{counterTest}))\mathrel{=}[\mskip1.5mu \mathrm{4},\mathrm{3},\mathrm{2},\mathrm{1}\mskip1.5mu]}\]
where \ensuremath{\Varid{nub}} is a function 
which removes
the duplicates from a
list.
\end{example}

\section{Related Work}
\label{sec:relatedwork}
From the categorical point of view, the theory of coalgebraic systems have been
thoroughly studied~\citep{RuttenT93, TuriP97, JacobsR12, Rutten00, Staton11} and
their connection with Structural Operational Semantics (SOS)~\citep{Plotkin04a}
have been thorougly expounded by ~\cite{Klin11}.

In functional programming, Turi and Plotkin's distributive laws~\citep{TuriP97}
have been proven to have wide range of applications: to prove the unique
fixed-point principle correct~\citep{HinzeJ11}; to define operational semantics
modularly~\citep{JaskelioffGH11}; and to prove equality of sorting
algorithms~\citep{HinzeJHWM12}. Hutton's work~\citep{Hutton98} aimed a
popularising the use of folds and unfolds for program semantics and the razor
has been used throughout the literature to explain key ideas of programming
languages design, semantics and compilers~\citep{Hutton98, Hutton21, BahrH15,
Hutton21}. In our work we take a slightly different approach. We tweak the razor
to fit into the idea that the \emph{denotational semantics is an algebra} on the
semantic domain defined as unfold and the \emph{operational semantics is a
coalgebra} defined as a fold on the syntax. Under this view the fold over the
denotational semantics is equal to the unfold over the operational semantics
and, by  using distributive laws, full abstraction falls out for free.  The
proofs of these facts have been thoroughly spelled out by~\cite{HinzeJ11} and \cite{HinzeJHWM12} which, on the other hand, do not discuss on the
full abstraction results.

\section{Discussion}
\label{sec:conclusions}
Recursion schemes were invented to ensure recursive definition are well-defined
mathematically. However, recursion schemes can go even further. While folds are
compositional interpretations, unfolds are fully abstract interpretations. In
particular, they can ``run'' a transition system (or an automata) and, for a state, construct
its trace such that equal traces correspond to operationally equivalent states.

When a transition system such as an operational semantics and a denotational
semantics arise from a distributive law, the fold over the denotational semantics
is equal to the unfold over the operational semantics.

\bibliography{../references}
\end{document}